\documentclass[11pt,preprint]{aastex}

\usepackage{graphicx}

\usepackage{natbib}

\slugcomment{Accepted by AJ}

\shorttitle{Comet C/2010 X1 Elenin}
\shortauthors{Li and Jewitt}

\begin{document}

\title{DISAPPEARANCE OF COMET C/2010 X1 (ELENIN): \\
 GONE WITH A WHIMPER, NOT A BANG}

\author{Jing Li\altaffilmark{1} and David Jewitt\altaffilmark{1,2} }
\affil{(1) Department of Earth, Planetary and Space Sciences,  University of California at Los Angeles}
\affil{(2) Department of Physics and Astronomy, University of California at Los Angeles}

\email{jli@igpp.ucla.edu}

\begin{abstract}

We examine the  rise and sudden demise of comet C/2010 X1 (Elenin) on its approach to perihelion.  Discovered inbound at 4.2 AU, this long-period comet was predicted to become very bright when near perihelion, at 0.48 AU on 2011 September 10.  Observations starting  2011 February (heliocentric distance $\sim$3.5 AU) indeed show the comet to brighten by about 11 magnitudes, with most of the increase occurring inside 1 AU from the Sun.  The peak brightness reached $m_R$ = 6 on UT 2011 August 12.95$\pm$0.50, when at $\sim$0.83 AU from the Sun.   Thereafter, the comet faded even as the heliocentric distance continued to decrease.  We find that most of the  surge in  brightness in mid-August resulted from dust particle forward-scattering, not from a sudden increase in the activity.   A much smaller ($\sim$3 magnitudes) brightening began on UT 2011 August 18$\pm$1 (heliocentric distance 0.74 AU), reached a maximum on UT 2011 August 30$\pm$1 (at 0.56 AU), and reflects the true break-up of the nucleus.  This second peak was matched by a change in the morphology from centrally condensed to diffuse. The estimated cross-section of the nucleus when at 1 AU inbound was $\sim$1 km$^2$, corresponding to an equal-area circle of radius 0.6 km.   Observations were taken after the second peak using the Canada-France-Hawaii 3.6-m telescope  to search for surviving fragments of the nucleus. None were found to a limiting red magnitude $r'$ = 24.4, corresponding to radii $\lesssim$40 m (red geometric albedo = 0.04 assumed).  The brightening, the progressive elongation of the debris cloud and the absence of a central condensation in data taken after UT 2011 August 30 are consistent with disintegration of the nucleus into a power law size distribution of fragments with index $q$ = 3.3$\pm$0.2 combined with the action of radiation pressure.  In such a distribution, the largest particles contain most of the mass while the smallest particles dominate the scattering cross-section and apparent brightness.  We speculate about physical processes that might cause nucleus disruption in a comet when still 0.7 AU from the Sun.   Tidal stresses and devolatilization of the nucleus by sublimation are both negligible at this distance.  However, the torque caused by mass loss, even at the very low rates measured in comet Elenin, is potentially large enough to be responsible by driving the nucleus to rotational instability.
\end{abstract}


\section{INTRODUCTION}

Comet C/2010 X1 (Elenin) (hereafter simply ``Elenin'') was discovered on UT 2010 December 10 as a $\sim$19.5  red magnitude object at heliocentric distance $r_H$ = 4.221 AU  \citep{2010CBET.2584....1E}.  The orbit was soon determined to be that of a long-period comet (semimajor axis, eccentricity and inclination were -7532 AU, 1.000064 and 1.84\degr, respectively) with perihelion  expected at $q$ = 0.482 AU on UT 2011 September 10, to be followed soon after by a close approach to Earth (minimum geocentric distance 0.234 AU on UT 2011 October 17).  Simple (but unphysical) power-law extrapolations of the apparent brightness from the discovery epoch gave rise  to predictions that Elenin would become a bright naked-eye comet  near and soon after perihelion.  These predictions at first seemed to be substantiated, as ground-based observers reported rapid brightening in early August  to peak  visual magnitudes near 8 \citep{2011CBET.2876....1S}.  However, Elenin was seen to be fainter on or about UT 2011 August 17 while the morphology was reported to change from centrally condensed to increasingly diffuse.  The comet was last recorded in unpublished observations by amateur astronomers as an extremely diffuse, elongated nebulosity on UT 2011 October 23, when outbound at $r_H$ = 1.06 AU.  In due course, comet Elenin disappeared.

Other promising long period comets (for example,  C/2012 S1 (ISON) and C/2011 W3 (Lovejoy)) have received considerable observational attention \citep[e.g.][]{2012ApJ...757..127S,2013ApJ...779L...3L,2014ApJ...782L..37K}.  The disintegration of these comets is relatively easily understood as a consequence of intense solar heating at their small perihelion distances (0.012 AU and 0.006 AU, respectively).  Comet Elenin, however, disintegrated even before reaching its relatively distant perihelion at 0.48 AU, under the action of incident solar fluxes $\gtrsim$5000 times smaller than those experienced by ISON and Lovejoy.  Elenin therefore offers a different perspective on the mechanism responsible for the destruction of a long period comet on its approach to perihelion.  Fading  is recognized as  important for understanding the steady-state population of long-period comets (specifically, it is needed to reconcile \citet{1950BAN....11...91O} dynamical model with the observed distribution of cometary orbital binding energies; \citet{1999Icar..137...84W}), but the mechanism responsible for fading is neither well-documented nor well-understood.  

Despite  widespread initial excitement and the spectacular physical development of Elenin, we are aware of only one relevant scientific publication in the refereed journals  (Korsun et al.~2012).  In the present paper, we present calibrated photometric measurements from a range of space-based and ground-based telescopes, documenting the demise of comet Elenin.  We use these measurements to interpret the mechanism behind its disappearance.

\section{OBSERVATIONS}
\begin{figure}
\epsscale{1.0}
\begin{center}
\includegraphics[width=0.9\textwidth]{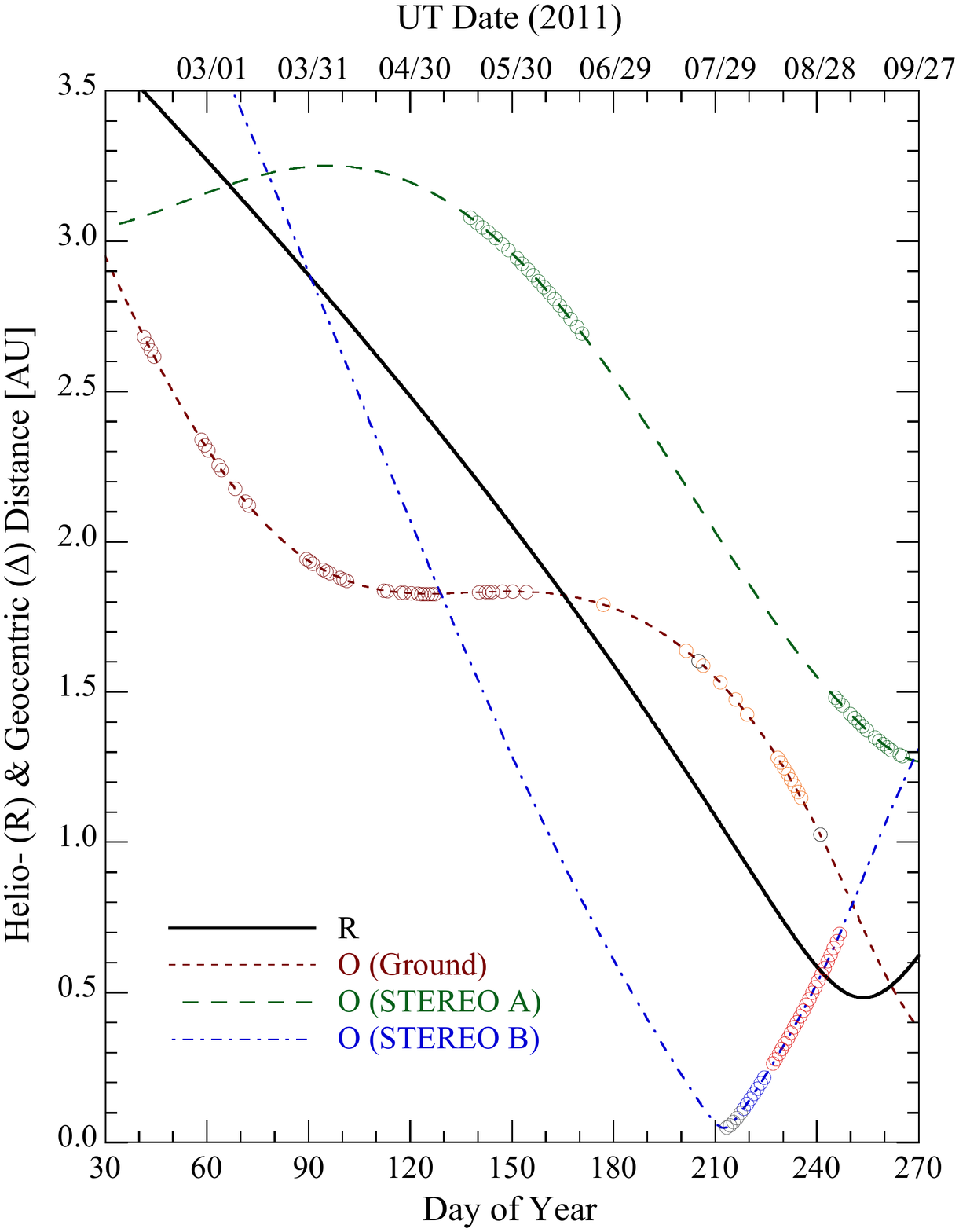}
\caption{Heliocentric (black) and geocentric distances (dotted brown, dashed green and dotted-dashed blue for Earth, STEREO A and STEREO B, respectively) in 2011. Differently colored circles represent observations from individual instruments as labeled in Fig. \ref{phase},  and  also explained in Table \ref{elenin}. \label{distance} }
\end{center} 
\end{figure}

\begin{figure}
\epsscale{1.0}
\begin{center}
\includegraphics[width=0.9\textwidth]{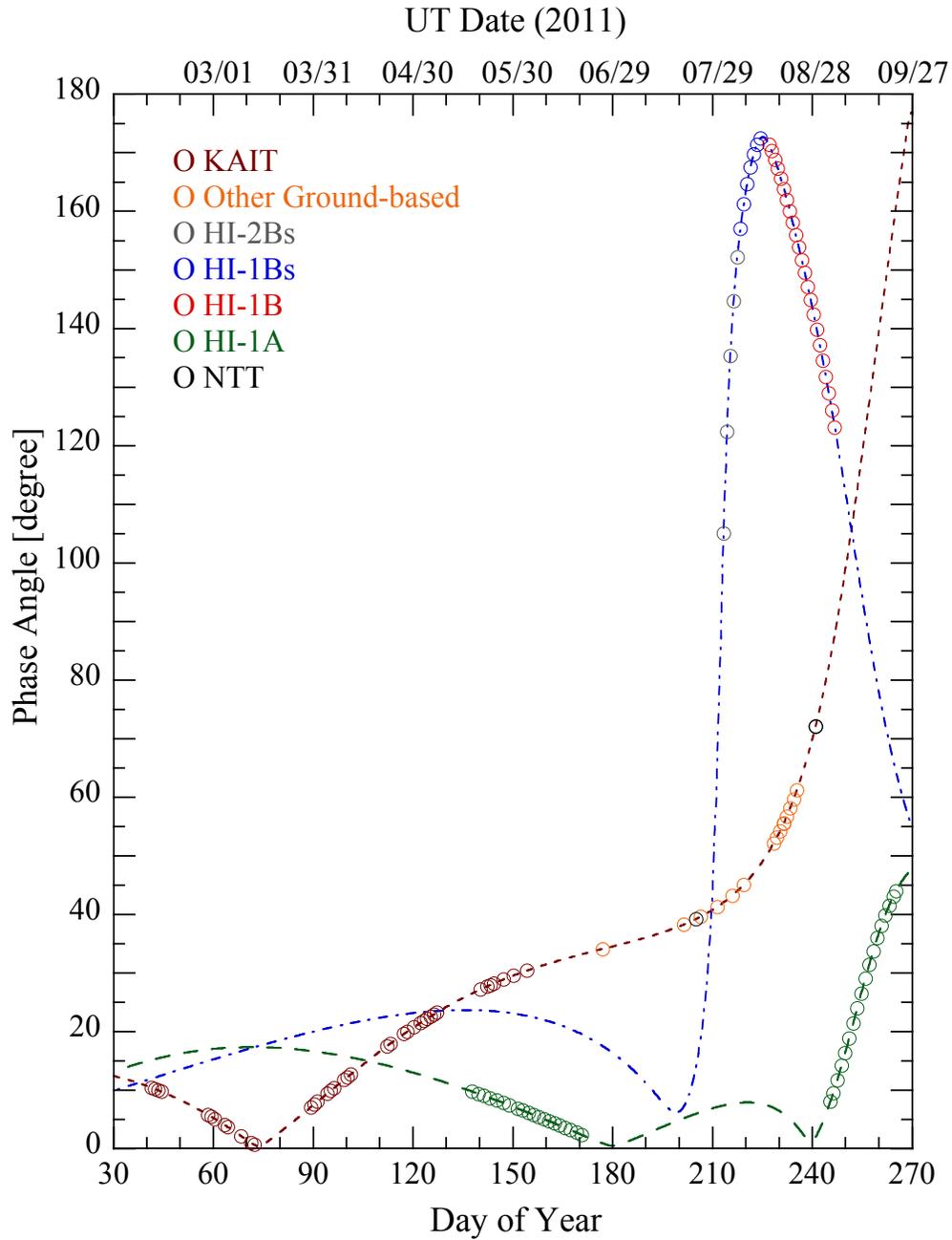}
\caption{Phase angle of the comet as viewed from different instruments, each indicated by differently colored circles. The dotted brown, dashed green and dotted-dashed blue curves represent the observations from Earth, STEREO A and STEREO B, respectively. \label{phase} }
\end{center} 
\end{figure}

We obtained comet observations using a mixture of ground-based and space-based telescopes between 2011 February and September in the inbound-lag  (see Table \ref{elenin}). Ground-based photometric monitoring observations of the approach to perihelion were obtained between UT 2011 February and 2011 June  using Berkeley's Katzman Automatic Imaging Telescope (KAIT).  For part of the period of interest, the solar elongation of comet Elenin as seen from Earth fell below 30\degr, effectively precluding ground-based observations.  However, the comet entered the fields of view of cameras onboard the Sun-observing spacecraft STEREO A and B between 2011 May 16 and September 22.  These spacecraft move in Earth-like orbits but are separated from the Earth in longitude by large angles \citep{2008SSRv..136...67H}, providing data from non-terrestrial perspectives and complementing the observations from Earth. In addition to their regular Sun-pointed observations, a set of targeted measurements was obtained using the STEREO B spacecraft. After the comet reached its peak-brightness in August and became faint in subsequent month prior to perihelion,  the Canada-France-Hawaii Telescope was used to conduct a sensitive search for nucleus fragments in October.  Lastly, archival measurements from the New Technology Telescope offered views of the comet at two epochs bracketing peak brightness at the end of July and the end of August.  Figures (\ref{distance}) and (\ref{phase}) show the comet geometry relative to the ground- and space-based telescopes between 2011 February and September as the comet approached the Sun. 

\subsection{Katzman Automatic Imaging Telescope (KAIT)}
The KAIT is a 0.76 meter telescope located at Lick Observatory, California, at an altitude of 1283 m \citep{2001ASPC..246..121F}. The telescope is mostly used for  supernova survey observations. We used KAIT to monitor the brightness evolution of Elenin between UT 2011 February 10 and June 3. The KAIT CCD  has  dimensions 512$\times$512 pixels (24 $\mu$m per pixel) with a field of view 6\arcmin.8$\times$6\arcmin.8 (at 0\arcsec.8 pixel$^{-1}$). The telescope tracks at sidereal rates without guiding, limiting the maximum usable exposures.  Comet Elenin was observed typically in a sequence of ten images, each of 30 s duration, for a total integration of 300 s per night in the R-band. The comet trailing does not substantially affect the photometry. The comet moved  with the maximum speed $\sim$54\arcsec hr$^{-1}$, leading to negligible image trailing ($\sim$ 0.5\arcsec~in 30 seconds). We used a photometry aperture 7.5 pixels (6\arcsec) in radius for the object and an annulus with inner and outer radii 9 and 12 pixels (7\arcsec.2~to 9\arcsec.6), respectively, to define the sky background. Photometric calibration was secured using Sun-like field stars, identified with the Aladin Sky Atlas and with R-magnitudes taken from the USNO-A2.0 catalog. The magnitude uncertainty, typically $\pm$0.1 magnitudes, was estimated from the scatter of 10 individual measurements of field stars. 

\subsection{STEREO Heliospheric Imagers }
The comet was detected during both regular and targeted observations by the Heliospheric Imagers (HI) on board the twin solar spacecraft STEREO A and B \citep{2009SoPh..254..387E}. On each spacecraft, HI is equipped with two cameras, HI-1 and HI-2, having small ($20\degr$) and large ($70\degr$) angular fields of view, respectively. The camera centers point away from the Sun by $14\degr$ (HI-1) and $53.7\degr$ (HI-2). The spectral passbands  of the HI-1 cameras cover wavelengths from 630 to 730 nm,  similar to the passband of the Johnson R-band filter.  However, the pre-launch calibration indicates about 25\% leak from the wavelengths at 300 - 450 nm \citep{2010SoPh..264..433B}. This may result a slight departure from the photometry from R-band. 

\begin{figure}
\epsscale{1.0}
\begin{center}
\includegraphics[width=0.8\textwidth,angle=-90]{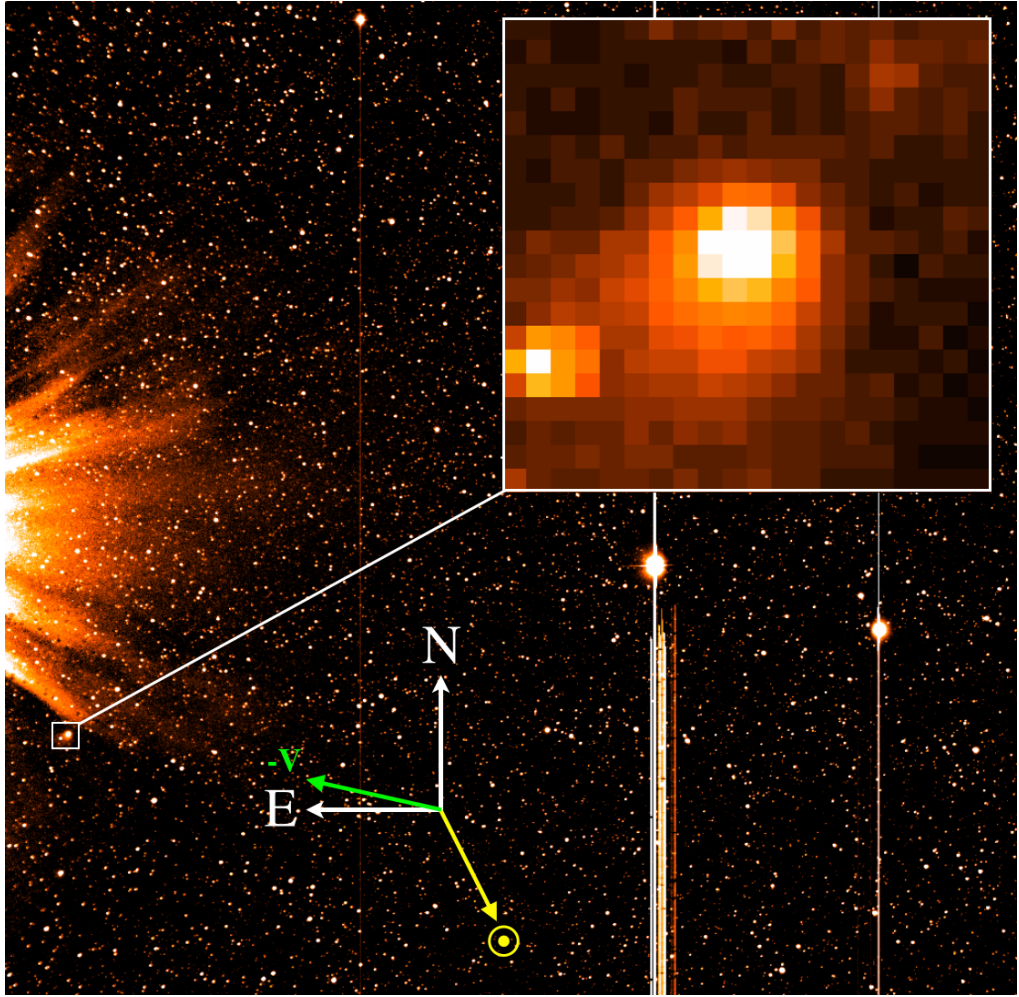}
\caption{Sample image from the STEREO B H1 camera taken UT 2011 August 15 at a solar elongation of 6.8\degr.  The full image shown is 20\degr~square and has been processed to remove the diffuse coronal background but not field stars, so that the density of these can be seen.   The inset shows a region 23\arcmin~square centered on comet Elenin.  Yellow and green arrows show the projected anti-solar direction and the negative of the heliocentric velocity vector.  The two brightest point sources are the planets Mercury and Jupiter.  \label{appearance} }
\end{center} 
\end{figure}

The STEREO HI data are publicly available and were downloaded from the UK Solar System Data Center. We retrieved  Level 1.0 data and proceeded to remove the background static coronal brightness using the same median procedure that we developed to study the asteroid Phaethon  \citep{2013AJ....145..154L}.  The comet should be in the field of view of the STEREO A HI-1 camera between May 16 and June 20, but we did not detect it because the comet brightness was below the detectable threshold of the camera during this time period. It was estimated that the comet had the brightest magnitude $\sim$15, while the camera threshold is about 12. The comet was observed, however, during the special maneuver of  STEREO B.  The wide-field camera, HI-2, recorded the comet  from August 01 to 05 while the narrow field camera, HI-1, recorded it from  August 06 to 12 (designated HI-1Bs). However, we found that the extremely poor angular resolution of the HI-2 camera rendered these data susceptible to excessive contamination by background coma and field objects.  As a result, we elected to not include HI-2 data in the present study. During the regular observations, the comet entered the HI-1/STEREO B field of view between August 14 and September 04 (designated HI-1B); and HI-1/STEREO A between September 01 to 22 (designated HI-1A). Figure (\ref{appearance}) shows the comet appearance in the camera on STEREO B. 

To increase the signal-to-noise ratio and remove background coronal structures and stars near the comet, we extracted the comet photometry from median images shifted to be centered on the comet.  During the targeted observations by STEREO B, the comet was observed for an hour a day. We obtained hourly median images of the comet and stars between August 01 and 12 (HI-1Bs).  In regular STEREO B and A observations, we made median images every 0.5 days from August 14 to September 2 with STEREO B (HI-1B), and every 2.4 days with STEREO A (HI-1A). The longer median durations were used to obtain better rejection of background stars as the comet was projected against the Milky Way  as seen from the latter spacecraft.

Equivalent median images were calculated for standard stars in the same way as for the comet. Field stars for photometric calibration were selected to be spectrally similar to the Sun and spatially close to the comet in the images. For the HI-1 camera, stars can be found in the SECCHI star catalog through an IDL routine scc\_get\_stars which is the part of SolarSoftWare (SSW) IDL package \citep{1998SoPh..182..497F}. The catalog contains more than 35,000 stars brighter than V = 10. It provides the V-magnitude of the stars; we converted from V to R by subtracting 0.36 (the color of the Sun) from the measurements. 

The STEREO HI-1 camera has a pixel size of 70\arcsec. We examined the cometary radial profile, and decided to use a circular aperture 3 pixels (210\arcsec) in radius, with the sky background determined in a concentric annulus having inner and outer radii of 6 and 9 pixels (7.0\arcmin~and 10.5\arcmin) for photometry. The sky annulus may be weakly contaminated by the comet tail but this contamination is preferable to the greatly enlarged uncertainties of measurement which are incurred by the use of larger annuli. The photometric uncertainties were  estimated based on the field star measurements. These uncertainties depend on the heavily under-sampled point-spread functions of the data  and on the residual background brightness. 

\subsection{The ESO New Technology Telescope}
\begin{figure}
\epsscale{0.5}
\begin{center}
\includegraphics[width=0.75\textwidth,angle=-90]{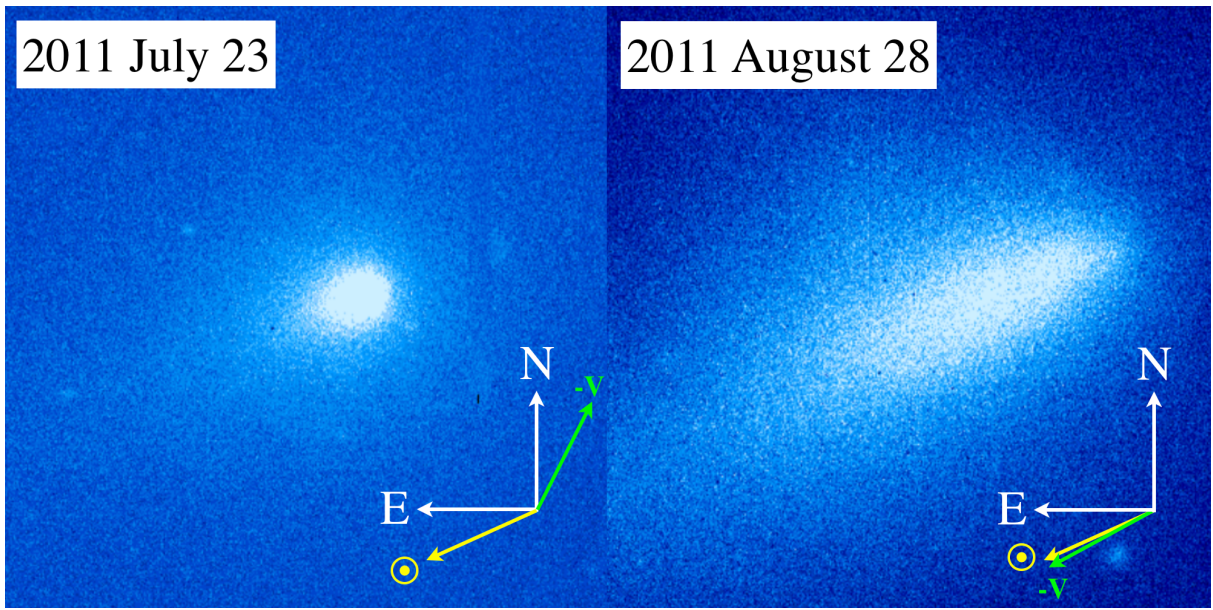}
\caption{Comet images taken by NTT on UT 2011 July 23 (left) and August 28 (right), showing a dramatic physical change. The image panels are 96\arcsec$\times$96\arcsec. Yellow and green arrows show, respectively, the position angles of the projected anti-solar direction and the negative of the projected heliocentric velocity vector. North and East are to the top and left, as marked. \label{ntt} }
\end{center} 
\end{figure}

We searched for Elenin observations in the archive data organized by the Canadian Astronomy Data Center.  ESO's New Technology Telescope (NTT), a 3.6m telescope, observed the comet on UT 2011 July 23 and August 28 (principal investigator Olivier Hainaut). The comet was observed with the  ESO Faint Object Spectrograph and Camera (EFOSC) \citep{1984Msngr..38....9B,2008Msngr.132...18S} on NTT  in the R-band. The EFOSC CCD field of view is 4.1\arcmin$\times$4.1\arcmin, while the image size is 2060$\times$2060 pixels. The images are binned 2$\times$2 pixels in the archival data giving an effective pixel scale 0.24\arcsec~(pixel)$^{-1}$. The telescope was tracking the comet. Figure \ref{ntt} shows the comet images from each date, revealing a dramatic change in the morphology between these two observations. 

\begin{figure}
\epsscale{1.0}
\begin{center}
\includegraphics[width=0.9\textwidth]{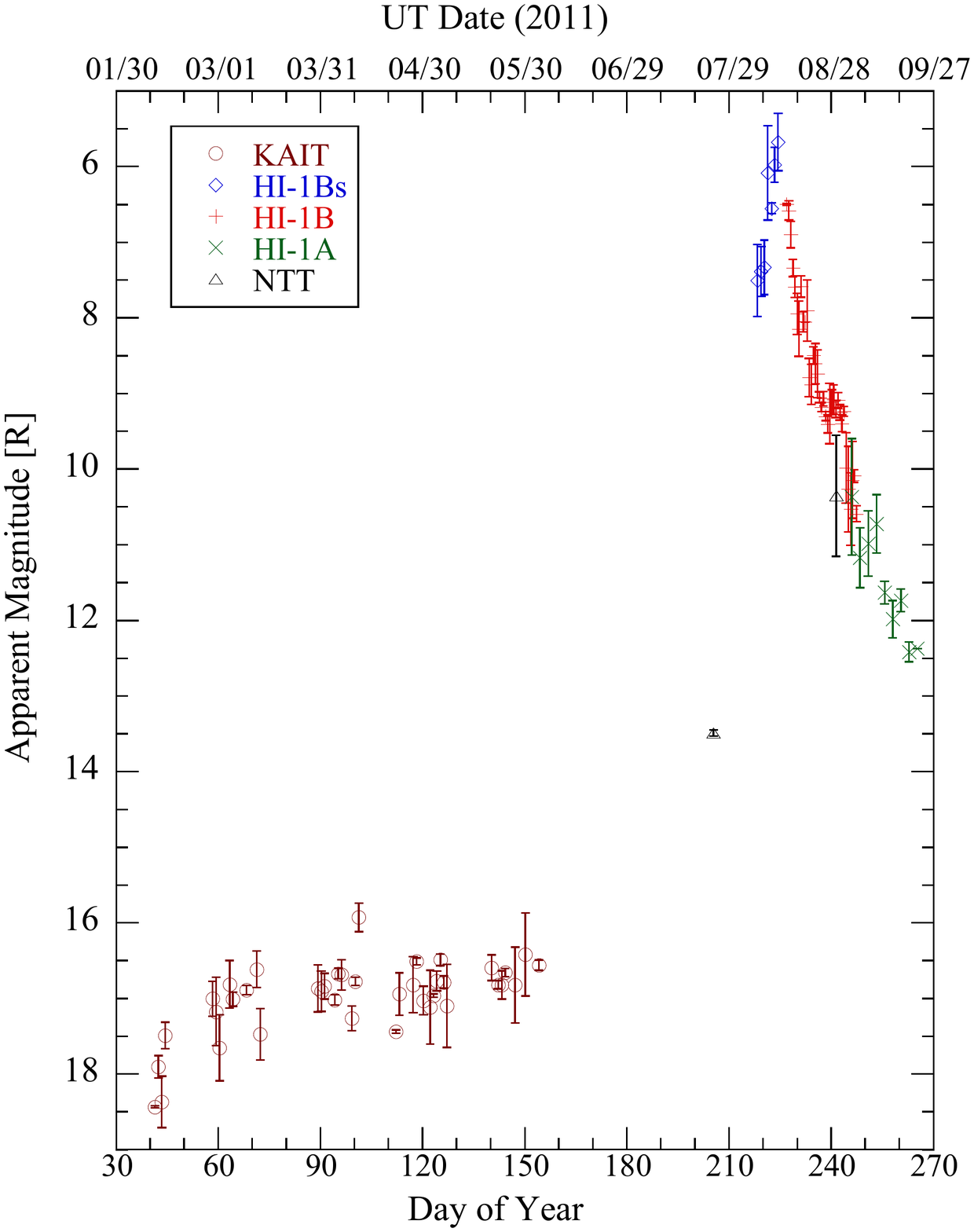}
\caption{Apparent magnitude of comet Elenin as a function of time expressed as Day of Year (bottom) and UT Date (top). Colored symbols represent the measurements from different instruments, KAIT: circles; HI-1Bs: diamonds; HI-1B: +; HI-1A: x; and NTT: triangles. The uncertainties are estimated from the scatter of repeated measurements at each date. The magnitudes presented in this figure and all other figures are R-magnitudes.\label{m_app} }
\end{center} 
\end{figure}

Because of the dramatic changes of the comet morphology, we used different photometry apertures for the two dates. The comet aperture was 140 pixels (33.6\arcsec), and the sky annulus was 150 to 200 pixel for the July data. The aperture was 350 pixels (84.0\arcsec), and the sky annulus was 350 to 400 pixels for the August data. We used all bright field stars in the field of view to  calibrate the photometry. The star photometry was extracted using apertures 15 pixels (3.6\arcsec) in radius with a concentric sky annulus from 15 to 20 pixels. Three images of the comet were taken on July 23  with exposure time 30 seconds within 3 minutes (UT 23:20- 23:23), and ten images were taken on August 28 within 11 minutes (UT 23:20-23:31), and exposure times ranged from 3 to 80 seconds. Assuming that the comet brightness did not change dramatically on such short timescales, we calculated average magnitudes of the comet observed in the July 23 and August 28. The uncertainties are estimated from the scatter of the repeated measurements.  In the July 23 data, the coma noticably over-spilled the photometry aperture.  However, experiments with different apertures indicate that the derived magnitude is within a few $\times$0.1 magnitudes of the ``total'' magnitude that would be obtained in an aperture of infinite radius.  The measurement on August 28 suffers more from aperture overspill, and should be regarded as a lower-limit to the brightness. The two data points from NTT are plotted in Figure (\ref{m_app}) along with apparent magnitudes from KAIT and STEREO A and B.

\subsection{Canada-France-Hawaii Telescope}
We used the 3.6 meter diameter Canada-France-Hawaii Telescope (CFHT) atop Mauna Kea, Hawaii, to examine Elenin on UT 2011 October 22.  The MegaCam prime focus imager was used with the r' filter to obtain a sequence of images at the expected position of Elenin.  MegaCam houses 36 CCDs each containing 2048$\times$4096 pixels of 0.187\arcsec~angular size, giving a total field of view of 1\degr $\times$ 1\degr.  The CFHT was tracked at non-sidereal rates during the observations so as to follow the motion of the comet against the stars.  Simultaneous guiding and non-sidereal tracking was not possible and therefore we limited the exposure duration to 120 s in order to minimize image smear due to open-loop tracking errors. The observing conditions were excellent, with 0.6\arcsec~FWHM seeing and photometric skies.  The observations were taken in queue-scheduled mode.

We obtained 22 images on UT 2011 Oct 22.  The CCDs within each image were combined using the SWarp software and compared visually on a computer.  No candidate objects showing the expected angular motion of Elenin were found, down to a limiting magnitude estimated to be $r'$ = 24.4.  At the expected position of Elenin, we did not detect a nucleus.

\subsection{Other Observations }
Between UT June 25.93 and August 23.39 2011, a number of amateur observers reported sightings of the comet \citep{2011CBET.2876....1S}. 
Visual magnitudes of active comets, especially when reported by different observers using different instruments, are very difficult to interpret and so we have not used them in the present analysis. However, the amateur observations do provide a qualitative reference for the comet brightness between  May and mid-July when data are  available from neither  KAIT nor STEREO. They show that the apparent brightness increased gradually during this time period, consistent with the slow brightening evident in earlier KAIT data, and reaching a peak on mid-August.

On July 30.153 and 30.249, M. Drahus and B. Yang detected HCN emission in comet Elenin using the James Clerk Maxwell Telescope \citep{2011CBET.2781....1D}. Assuming an isotropic production of gas at velocity 0.5 km s$^{-1}$ and a Boltzmann distribution of energy levels at 50 K, the derived HCN production rate is $1.5\times 10^{25}$ molecule s$^{-1}$. They noted that the HCN production is comparable to the mean level measured in comet 103P/Hartley at the same (1.07 AU) heliocentric distance in late 2010 \citep{2011ApJ...734L...4D}. At this time the  estimated visual magnitude was $\sim$9.7  (\citet{2011CBET.2801....2G}). During the same period, the comet's trajectory was nearly on the same plane as the Earth's orbit and its minimum distance to STEREO B was  only 0.05 AU. The spacecraft was literally within reach of the comet tail; the PLASTIC instrument on STEREO B directly detected suprathermal pickup H$^+$ and He$^+$ and singly charged water group ions produced by outgassing from the comet \citep{2013EGUGA..1512004K}.  While serving to confirm that the comet was a significant source of water at the end of July, it is not practical to use the plasma detections alone to estimate the production rate.

The comet was too close to the Sun to be seen from the Earth in 2011 September. A search on UT October 9.5 and 10.6, using the Faulks Telescope North having a field of view 10\arcmin$\times$10\arcmin, met with no success \citep{2011CBET.2876....2S}. On UT Oct. 21.38 and 21.48, the same observers reported a detection of a large and diffuse cloud 4\arcmin.3 from the predicted comet position in the east-northeast direction (position angle 77\arcdeg). This observation was made with the GRAS 0.1-m f/5 APO refractor at the Mayhill station in New Mexico (field-of-view 3.9\arcdeg$\times$2.6\arcdeg; scale 3.5\arcsec~ (pixel)$^{-1}$). The large blob moved with the predicted comet motion in the sky. On Oct. 23.4. with the same telescope, \citet{2011CBET.2876....1S} confirmed the ``extremely faint and diffused blob" with an extended size 1\degr.5$\times$10\arcmin.0.  This blob appears to be the expanded and diluted remnant of the post-break-up dust cloud imaged at NTT on UT 2011 August 28 (Figure \ref{ntt}).

On November 8.12, the amateur astronomer Bernard H\"{a}usler, in Germany, took images at the predicted location of comet Elenin. He used a 30-cm Schmidt-Cassegrain with a field of view 29\arcmin.7$\times$20\arcmin.0 and the pixel scale 1\arcsec.63. The comet was not detected. 

\section{RESULTS}
Our observations make it clear that Elenin did not conform to the simple power-law brightening model used to make optimistic predictions in 2010 and 2011.  In this section, we examine the brightness variations and attempt to use them to diagnose physical processes occurring at the comet.  

\subsection{The Aperture and Distance Corrections}
The apparent magnitude of the comet  is shown as a function of time in Figure \ref{m_app}. The magnitudes were determined using different photometry apertures and with the comet  at different distances from the telescopes. Therefore, the apparent magnitudes are strictly not comparable, since they sample different volumes of space around the nucleus and different amounts of encircled dust.  To take account of this, we first make an ``aperture correction'', in which the apparent magnitudes are scaled to the magnitudes which would be observed if each telescope could sample a fixed volume (represented by a sphere of fixed radius) around the nucleus.  For this purpose, we arbitrarily scale the photometry to estimate the magnitude that would have been obtained if the comet were observed using a 100\arcsec~aperture from a distance of 1 AU (corresponding to a linear radius of 73,000 km).  

To make the aperture correction, we need to know how the coma surface brightness, $\Sigma(\phi)$, varies as a function of the photometry aperture radius, $\phi$.   Isotropic coma expansion at constant speed would give $\Sigma(\phi) \propto \phi^{-s}$ with $s$ = 1.0, while acceleration of the dust under a constant radiation pressure gives a steeper gradient, $s$ = 1.5  \citep{1987ApJ...317..992J}.  In the former case, the encircled dust coma cross-section (and, therefore, the brightness) grows with radius as $\int 2 \pi \phi \Sigma(\phi) d\phi \propto \phi$, and in the latter as $\propto \phi^{1/2}$.  The resulting aperture correction is given by 

\begin{equation}
\Delta m = -2.5(2-s) \log_{10}\left(\frac{100}{\phi\Delta}\right)
\label{aperture}
\end{equation}

\noindent where $\phi$ is the photometry aperture radius in arcseconds, $\Delta$ is the observer to comet distance in AU and $s \ne$ 2.  

The surface brightness profile of the coma was measured by \citet{2012AstBu..67..414K} on UT 2011 March 28.  They found a power-law relation with $s$ = 1.56$\pm$0.01, close to the value expected for a radiation-pressure swept coma.  Unfortunately, the coma was not well resolved in the KAIT data owing to the surface brightness sensitivity being limited by the short exposures and the brightness of the sky on Mount Hamilton. Surface brightness measurements in STEREO data are also limited by the large pixel size (70\arcsec~pixel$^{-1}$) and the bright, near-Sun background of coronal and scattered photospheric light.  We assume that the KAIT and STEREO observations of Elenin can be well-represented by $s$ = 1.5, although we strictly possess evidence validating this assumption only around the time of the \citet{2012AstBu..67..414K} observation.  The coma is fully resolved in the  NTT data, where measurements within circular, concentric apertures show that the inner surface brightness on UT 2011 July 23 is best described by $s$ = 1.0.  The NTT data from UT 2011 August 28 are insufficient to permit a meaningful determination of $s$, owing to the low surface brightness of the extended debris trail on this date.  To be consistent with July data by NTT, we take $s=1.0$ for the August NTT comet image.

\begin{figure}
\epsscale{1.0}
\begin{center}
\includegraphics[width=1.0\textwidth]{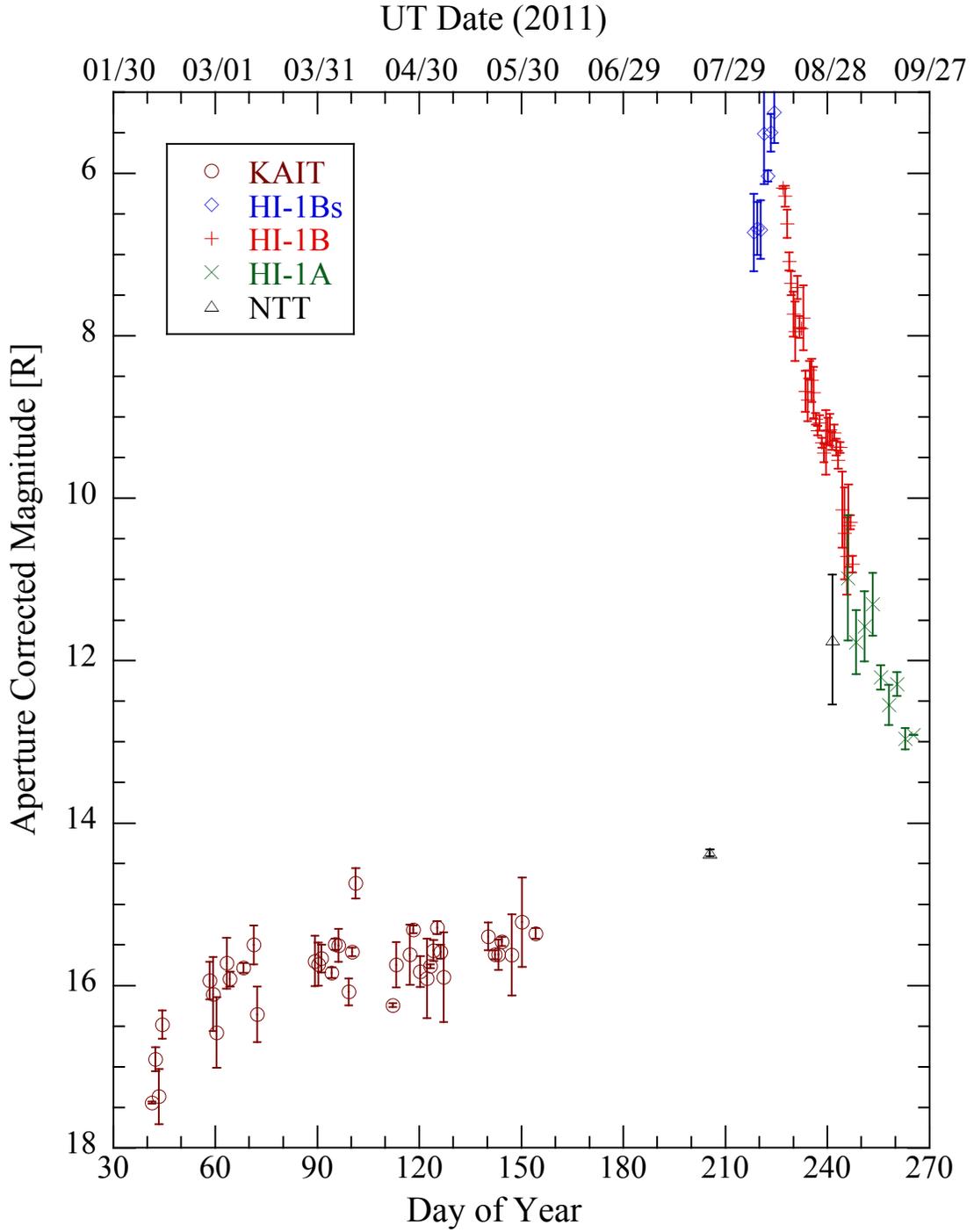}
\caption{Comet magnitudes after the application of the aperture correction (c.f.~Equation \ref{aperture}). \label{mr1a_100} }
\end{center} 
\end{figure}

Evidence that the aperture correction is at least approximately correct is provided by Figure \ref{mr1a_100}, in which data from different apertures and different telescopes are seen to overlap within the uncertainties of measurement.  The KAIT,  HI 1A, 1B, 1Bs and NTT data are all broadly consistent in the figure even though they employed quite different photometric apertures and were taken at different telescope-to-comet-distances.  A grave error in the aperture correction would cause a mis-match between datasets taken from different observatories.  Even after the aperture correction, an error is incurred because distant parts of the comet are excluded by the use of a finite aperture including the sky contamination by the coma.  This error is modest (a few tenths of a magnitude or less) until the middle of August, when the morphology changes from centrally-condensed to diffuse.  A compensating factor is that the post-peak photometry used very large apertures (210\arcsec~in radius) but it is still true that the post-peak measurements should properly be regarded as setting a lower limit to the integrated brightness of the whole coma.  The use of still larger apertures is precluded by the increased sky noise introduced by the bright background of coronal and scattered light, as well as by imperfectly removed field objects.

We used the aperture correction and the inverse square law to calculate the absolute magnitude of the comet at a given phase angle (i.e.~the magnitude corrected to $r_H$= $\Delta$ = 1 AU) from

\begin{equation}
m(1,1,\alpha) = m(r_H,\Delta, \alpha) - \Delta m- 5 \log_{10}(r_H\Delta)
\label{magnitude}
\end{equation}

\noindent where $m(r_H, \Delta, \alpha)$ is the apparent magnitude and $\Delta m$ is from Equation (\ref{aperture}). The aperture and distance-corrected magnitudes of the comet are plotted in Figure (\ref{fm11a}).  Variations in the photometry shown there result from a combination of the effects of real changes in the scattering cross-section of the dust within the scaled photometry aperture and changes due to the angle-dependence of the efficiency of the dust scattering (the so-called ``phase function'').

\subsection{The Phase Function}
As the comet approached the Sun, it entered a strongly forward-scattering geometry, reaching the maximum phase angle $\alpha = 172.6\degr$ as observed from STEREO B (Figure \ref{phase}). While Sun-grazing comets are often observed by LASCO/SOHO at moderate to  large phase angles \citep{2004A&A...427..755G}, ground-based telescopes are rarely able to point near the Sun and so large phase angle observations are rare. The current record for maximum phase angle is held by P/2003 T12 observed at $\alpha$ = 177.6\degr~(Hui 2013).    Observations of comets at large phase angles show that the dust is strongly forward-scattering \citep{1976Sci...194.1051N,1982come.coll..323N, 2007ICQ....29...39M}. 

\begin{figure}
\epsscale{1.0}
\begin{center}
\includegraphics[width=0.95\textwidth]{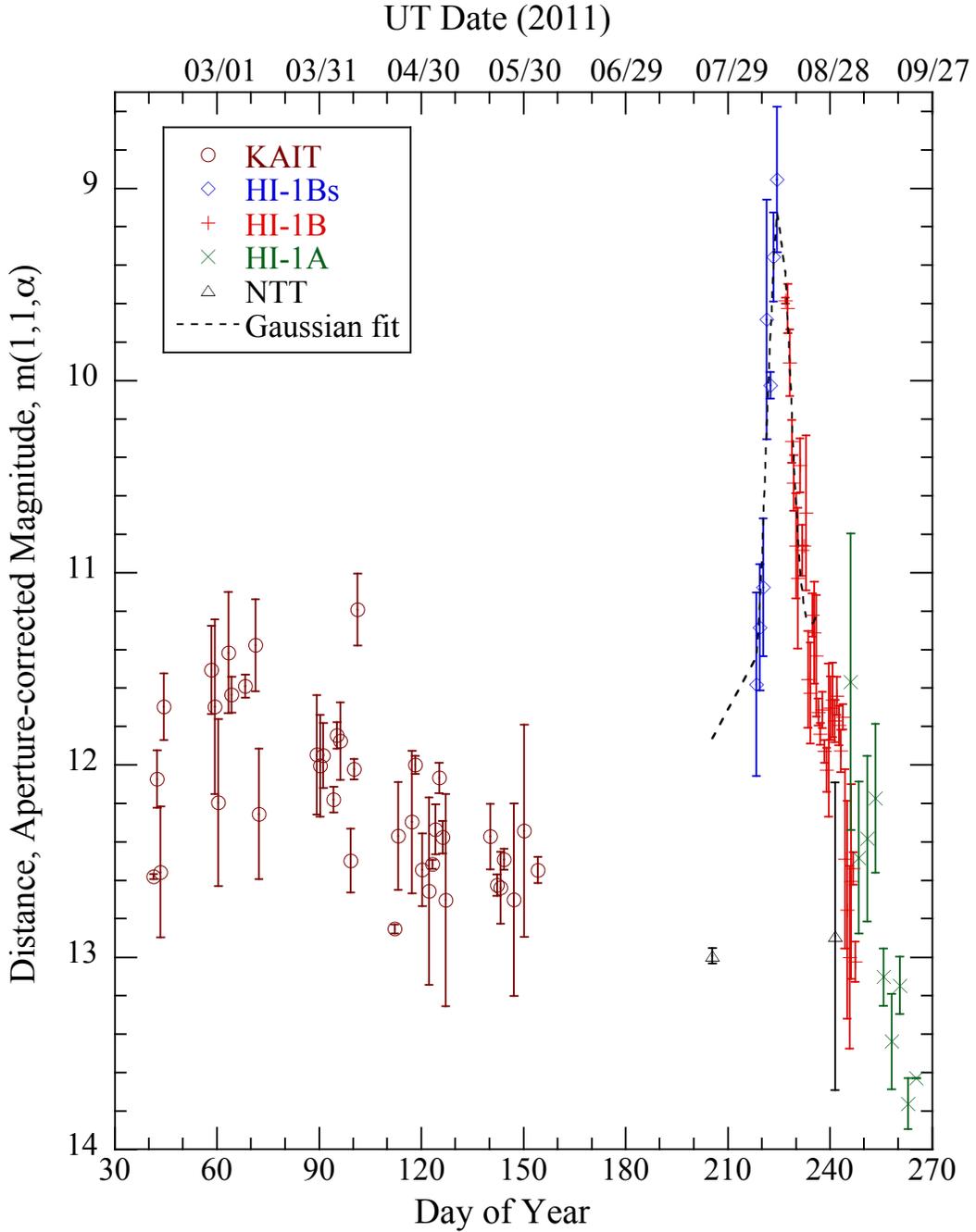}
\caption{Comet magnitudes after the application of both the aperture correction and the inverse square law correction (c.f.~Equation \ref{magnitude}). The peak of the light curve is fitted with a Gaussian profile having a center on UT 2011 August 12.95$\pm$0.5 (dotted curve). The full width at the half maximum of the Gaussian profile is 8 days. \label{fm11a} }
\end{center} 
\end{figure}

We first fitted a Gaussian function to the portion of the aperture- and distance-corrected light curve near the photometric maximum,  between UT August 06.3 and 18.5. The best fit gives the date of peak brightness as UT 2011 August 12.95. The uncertainty of the peak date is about 0.5 days estimated by a visual examination. The maximum phase angle of the comet, $\alpha = 172\degr.6$,  occurred on UT 2011 August 13 as viewed from STEREO B.  This coincidence between the dates of maximum brightness and maximum phase angle is a strong evidence that the main brightness peak in Figures (\ref{m_app}), (\ref{mr1a_100}) and (\ref{fm11a}) is caused by the cometary dust phase function and the forward-scattering observational geometry, not by a true outburst in the comet. \citet{2011CBET.2876....2S} inferred breakup on UT 2011 August 16$\pm$4, consistent with our estimate.  

The comet brightened by $\sim$3 magnitudes due to the phase angle effect (Figure \ref{fm11a}). However, the light curve is not completely symmetric about the maximum phase angle, presumably because of the sudden release of debris caused by the breakup of the nucleus (as witnessed by the dramatic change in the morphology of the comet near this time - see Figure (\ref{ntt})). To derive a phase function, we use the portion of the photometry data taken by KAIT, HI-1Bs and NTT before the maximum phase angle of the comet. This is the time period in which we can reasonably assume that there was no outburst based on the observations.

Following \citet{2007ICQ....29...39M}, we adopt a compound Henyey-Greenstein (HG) function \citep{1941ApJ....93...70H} to represent the  comet dust.
The HG function has no physical basis,  but is useful to provide an empirical fit to the phase functions of dust:

\begin{equation}
\Phi(\theta)=\frac{A(1-g^2)}{4\pi}\left[\frac{1}{(1+g^2-2g\cos\theta)^{3/2}}\right]
\end{equation}

\noindent where $\Phi$ is the normalized flux; $\theta = 1 - \alpha$ is the scattering angle (the deviation of the ray from the forward direction),  $A$ is the dust grain albedo, and $-1 \le g \le+1$ is the asymmetry factor, defined by $g=\int_0^\pi 2\pi\Phi(\theta)\cos\theta  \sin\theta d \theta$. Physically, $g$ is the scattered intensity weighted by $\cos\theta$ averaged over the entire solid angle. The HG function  is normalized such that the integral over $4 \pi$ steradians is unity: $\int_0^{2\pi}[\int_0^\pi\Phi(\theta)\sin\theta d \theta] d\phi=1$. Large positive (negative) values of $g$ represent strong forward (back) scattering. Isotropic scattering has $g$ = 0. We fitted the aperture and distance-corrected flux ($10^{-0.4\times m(1,1,\alpha)}$) observed by KAIT, HI-1Bs and NTT before the comet phase angle reached the maximum to a combined forward and back scattering HG function

\begin{equation}
f_{HG}(\alpha)=F_f\Phi(g_f)+F_b\Phi(g_b).
\label{fhg}
\end{equation}

\begin{figure}
\epsscale{1.0}
\begin{center}
\includegraphics[width=0.95\textwidth]{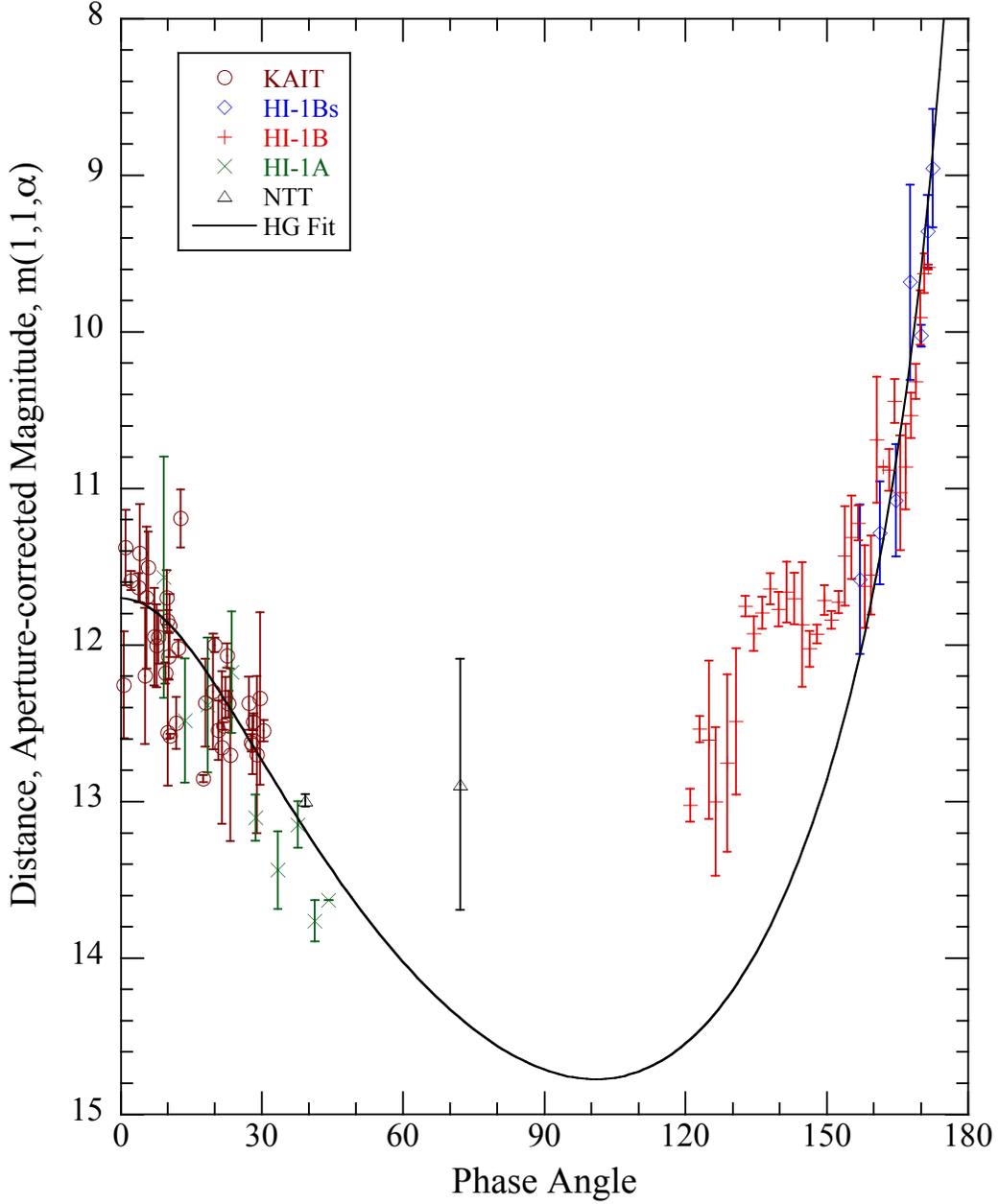}
\caption{Same as Figure (\ref{fm11a}) but plotted versus the phase angle. The pre-peak magnitudes (brown, black and blue circles) are fitted by a combined Henyey-Greenstein function (Eq. \ref{fhg}) to represent both forward and backward scattering (solid line). \label{m11a_hg} }
\end{center} 
\end{figure}

In Equation (\ref{fhg}), four free parameters are to be determined:  $g_f>0$ and $g_b<0$ are the forward and backward scattering $g$-parameters; $F_f$  and $F_b$ are scaling constants to the distance-corrected flux for the forward and back scattering components.  A $\chi^2$ fit yields values for $g_f=0.926$ and $g_b=-0.584$;  $F_f=8.075\times 10^{-4}$ and $F_b=2.850\times10^{-4}$ with A=0.1. The large $g_f$ is a result of forward scattering by particles much larger than the wavelength of observation.  For comparison, (mostly sub-micron) dust in the interstellar medium has a more modest $g_f \sim$ 0.5 to 0.6 \citep{2003ApJ...598.1017D}.  The fitted HG function is plotted in Fig (\ref{m11a_hg}). \citet{2007ICQ....29...39M} re-normalized the compound Henyey-Greenstein function at the phase angle $\alpha=90\degr$ as the dust scattering function. The ``partitioning coefficient'' $k$ in his work is equal to the ratio $F_f/(F_f+F_b$) in the current work. We obtain $k=0.74$. This is smaller than the value 0.95 used by \citet{2007ICQ....29...39M} indicating that comets do not share a single phase function, presumably because their dust properties differ. Based on the best-fit HG function, the brightness of the comet that would have been observed in pure forward-scattering ($\alpha=180\degr$) is 120 times that of the backscattered brightness ($\alpha=0\degr$).  Our best estimate of the pre-perihelion absolute magnitude (i.e.~corrected to $r_H = \Delta = 1$ AU and to $\alpha$ = 0\degr) is $m_R(1,1,0)$ = 11.7.

\begin{figure}
\epsscale{1.0}
\begin{center}
\includegraphics[width=0.95\textwidth]{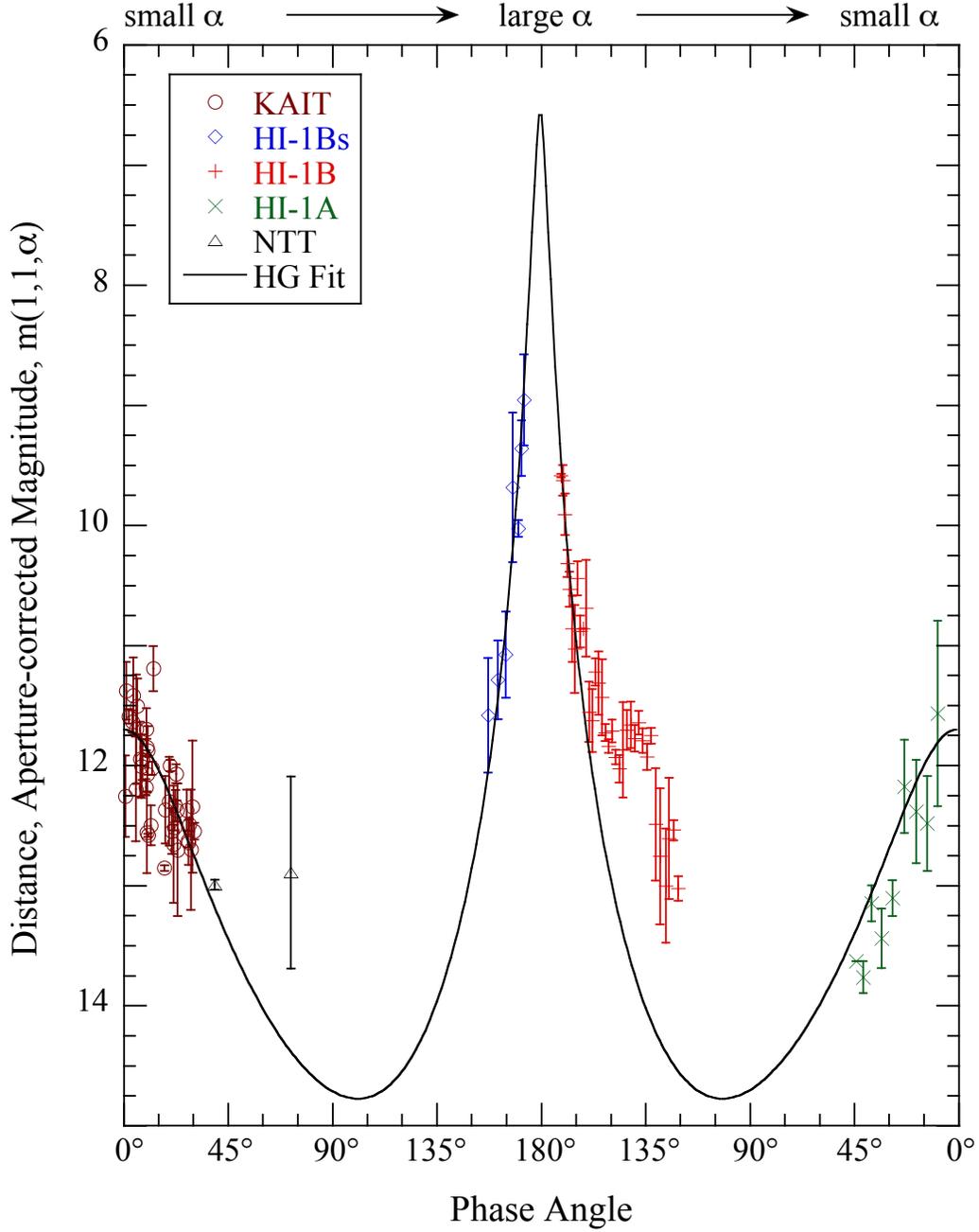}
\caption{Same as Figure (\ref{m11a_hg}) but separating phase angles before and after the peak.  The thick black curve is the fitted Henyey-Greenstein function. \label{m11a_hg2} 
}
\end{center} 
\end{figure}

Figure (\ref{m11a_hg2}) shows  how the comet brightness varied with the phase angle, this time plotted so as to distinguish the pre-phase-angle-maximum and post-phase-angle-maximum legs of the orbit.   Post-phase-angle-maximum brightening relative to the phase curve is evident, with a local maximum near $\alpha \sim$ 140\degr.

The relation between the absolute magnitude and the effective scattering cross section of the dust, $C_e$, (in km$^2$) is by \citet{1996LPI....27..493H} 

\begin{equation}
C_e = \frac{1.5\times 10^6}{p_V}10^{-(0.4 m_V(1,1,0))}
\label{inversesq}
\end{equation} 

\noindent  where $p_V$ is the geometric albedo, which we take to be 0.04 (representative of the albedos measured for the nuclei of comets).  

\begin{figure}
\epsscale{1.0}
\begin{center}
\includegraphics[width=0.95\textwidth]{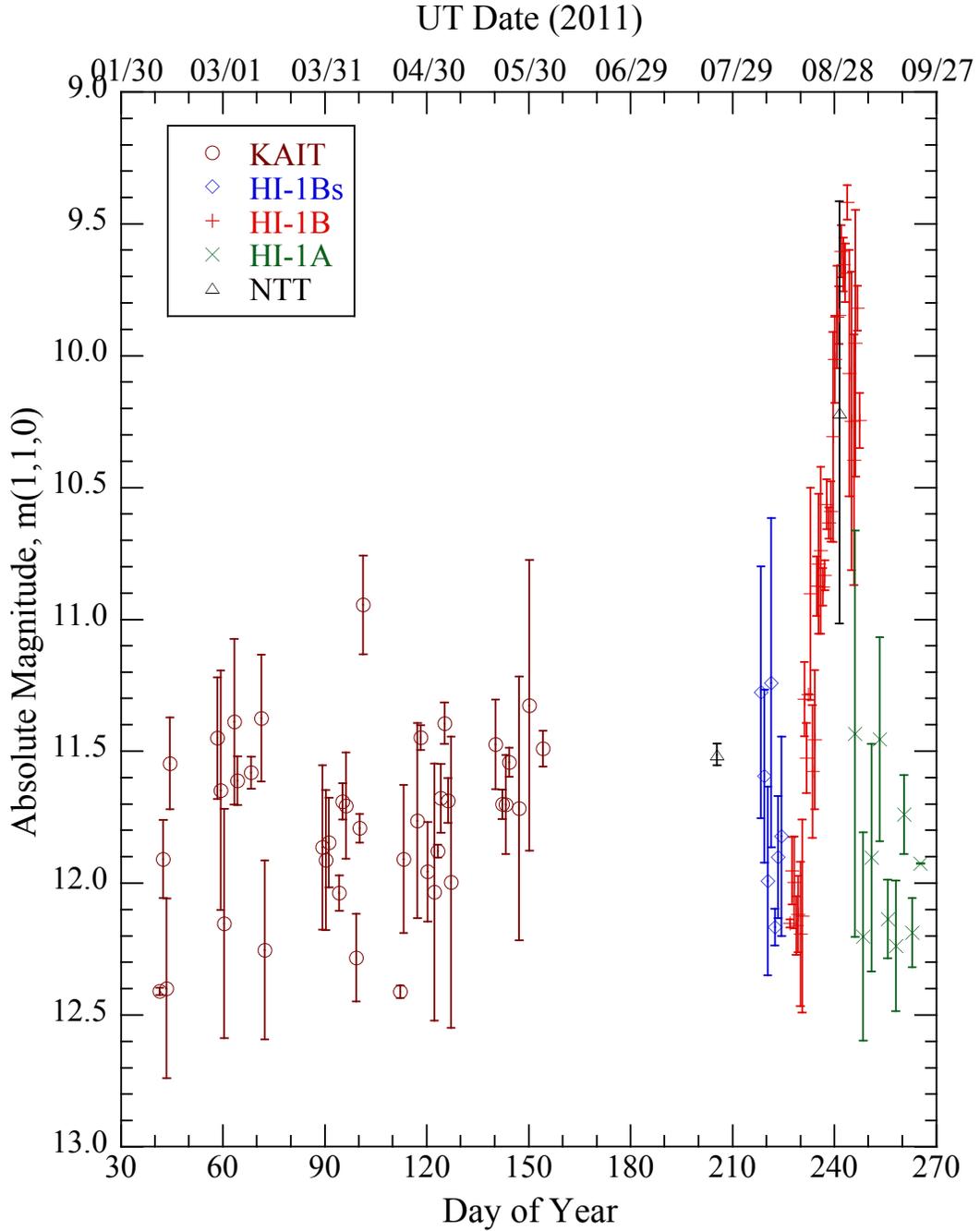}
\caption{Absolute magnitudes of Elenin, including aperture correction,  inverse square law correction and phase function correction. The measurements are scaled such that the comet brightness at $r_H=\Delta$=1 and $\alpha=0\degr$ is 11.7 magnitude, equal to the best-fit value from the Henyey-Greenstein fits. \label{m110} }
\end{center} 
\end{figure}

The absolute magnitude (i.e.~with both the aperture correction and the distance correction taken into account) is shown as a function of time in Figure (\ref{m110}). The absolute magnitude remained unchanged at $m_R(1,1,0) \sim$ 11.7$\pm$0.5 from UT 2011 February  through August (40 $\le DOY \le$ 230), in data from the KAIT, NTT and HI-1Bs. Substituting $m_R(1,1,0)$ = 11.7 and including a small correction for the color (we assume V-R = 0.36), we find $C_e$ = 570$\pm$300 km$^2$

\begin{figure}
\epsscale{1.0}
\begin{center}
\includegraphics[width=0.95\textwidth]{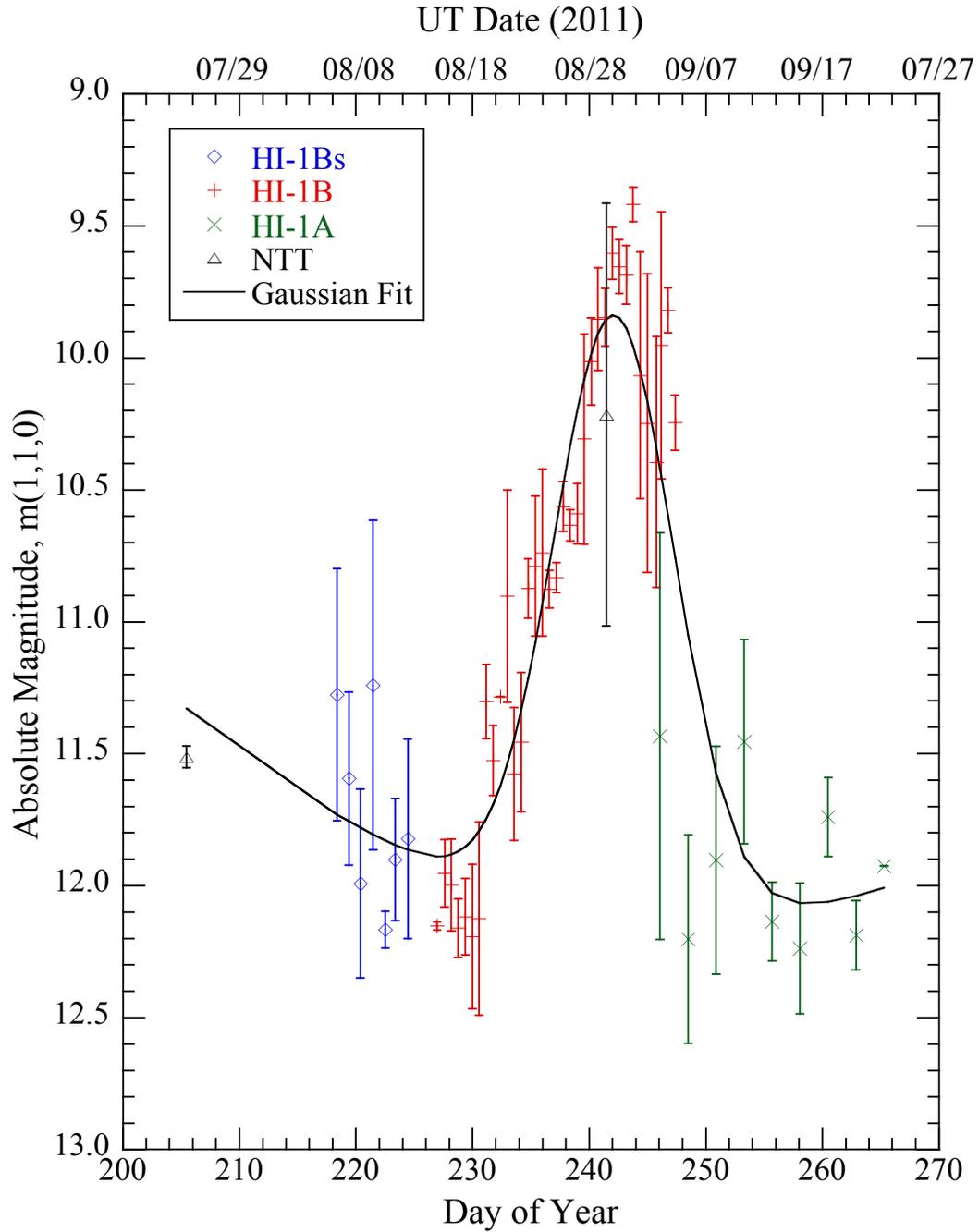}
\caption{Enlargement of Figure (\ref{m110}) near the time of the break-up event. A Gaussian fit gives peak brightness on UT 2011 August 30.1$\pm$0.5 and the peak brightness reached 2.6 magnitude from the base. The full width at half maximum of the Gaussian profile is 13.8 days.  \label{m110p} }
\end{center} 
\end{figure}

In Figure (\ref{m110}) the main forward-scattering peak  centered on August 13  is gone, but a separate and later peak appears at the end of August. Figure (\ref{m110p}) is an enlargement of Figure (\ref{m110}) showing the STEREO and NTT data recording this  peak. We again fitted this second peak with a Gaussian profile, finding a center on UT 2011 August 30.14 at $m_R(1,1,0) = 9.8\pm0.5$ and a full width at half maximum $\sim$12.2 days. This second peak appears unrelated to the phase angle and, instead, we interpret it as a real increase in the scattering cross-section of Elenin.  Substituting into Equation (\ref{inversesq}), we obtain $C_e$ = 4300$\pm$900 km$^2$ at the maximum of this second peak.  The rise towards maximum brightness begins near DOY 230$\pm$1 (UT 2011 August 18$\pm$1), which we take as the start of the disintegration of the nucleus of Elenin.

\subsection{Gas Production}
Spectral detections of CN and C$_3$ molecules on UT 2011 March 28 ($r_H$= 2.92 AU) gave production rates $Q_{CN}$ =  1.4$\times$10$^{24}$ s$^{-1}$ and $Q_{C_3}$ = 4.2$\times$10$^{23}$ s$^{-1}$, respectively \citep{2012AstBu..67..414K}. A very modest production rate of hydrogen cyanide, $Q_{HCN}$ = 1.5$\times$10$^{25}$ s$^{-1}$, was determined from pre-perihelion submillimeter spectroscopy on UT 2011 July 30 at $r_H$= 1.07 AU \citep{2011CBET.2781....1D}.  Both CN and HCN are trace species in the comae of comets.  We scaled their production rates to that of water using the nominal ratio $Q_{OH}/Q_{CN}$ = 320 \citep{1995Icar..118..223A}, finding $Q_{H_2O} \sim$ 4.5$\times$10$^{26}$ s$^{-1}$ (12 kg s$^{-1}$) at $r_H$= 2.92 AU and $Q_{H_2O} \sim$ 4.8$\times$10$^{27}$ s$^{-1}$ (128 kg s$^{-1}$) at $r_H$= 1.07 AU.  \citet{2011IAUC.9232....3L} failed to detect hydroxyl (OH) emission close to perihelion ($r_H$ = 0.49 AU) on UT 2011 September 07, setting an upper limit to the production rate of water $Q_{H_2O} \le$ 3$\times$10$^{27}$ s$^{-1}$ (80 kg s$^{-1}$).  These modest production rates are consistent with sublimation from very limited areas of exposed ice, as we next discuss. On 2011 June 28 ($r_H=1.6$ AU), Schleicher (private communication, 2015 February 6) obtained the $Q_{OH}=10^{27.8}$ s$^{-1}$ with a $\pm40$\% uncertainty, and $Q_{CN}=10^{25.04}$ s$^{-1}$. The resulting ratio,  $Q_{OH}/Q_{CN}$ = 575$\pm 230$, is consistent with the \citet{1995Icar..118..223A}  value of 320 that we used to estimate the water production rate. 


We solved the energy balance equation for water ice, including energy gained and lost by radiation and energy consumed as latent heat in the sublimation of water.   This equation is written

\begin{equation}
\frac{F_{\odot} (1 - A)}{r_H^2} = \chi \left[\epsilon \sigma T^4  + L(T) F_s\right]
\label{energy}
\end{equation}

\noindent in which $F_{\odot}$ = 1360 W m$^{-2}$ is the Solar constant, $A$ is the Bond albedo, $\epsilon$ is the emissivity, $\sigma$ is the Stefan-Boltzmann constant, $L(T)$ is the latent heat of sublimation for water ice and $F_s$ is the equilibrium sublimation flux (kg m$^{-2}$ s$^{-1}$).  The term on the left hand side describes the absorbed solar power while the two terms on the right describe radiation from the surface into space and power used to sublimate ice at rate $F_s$.  Another, much smaller term representing conduction into the interior has been neglected.  Dimensionless parameter $\chi$ represents the ratio of the absorbing area on the nucleus to the area from which absorbed heat is radiated.  Limiting values range from $\chi$ = 1 (a subsolar ice patch on a non-rotating nucleus) to $\chi$ = 4 (a spherical, isothermal nucleus in which the Sun's heat is absorbed on $\pi r^2$ and radiated from 4$\pi r^2$).  However, measurements of thermal radiation from cometary nuclei show that night-side radiation is negligible, as a result of the low diffusivity of the surface layers (Fernandez et al.~2013).  Likewise, in-situ imaging from spacecraft shows that the bulk of the outgassing from active comets occurs on the hot day side.  For a spherical nucleus, day-side only emission corresponds to $\chi$ = 2.  Accordingly, we use $\chi$ = 1 and $\chi$ = 2 to bracket the highest and lowest plausible temperatures and sublimation rates on the nucleus, respectively.   For the other parameters we take $A$ = 0.04, $\epsilon$ = 0.9, $\sigma$ = 5.67$\times$10$^{-8}$ W m$^{-2}$ K$^{-4}$, while $L(T)$ is obtained from \citet{1926_book}. 

The effective area of sublimating ice needed to supply gas at rate $dM/dt$ is given by

\begin{equation}
A_s = \frac{dM/dt}{F_s}.
\label{A_s}
\end{equation}

\begin{figure}
\epsscale{1.0}
\begin{center}
\includegraphics[width=0.8\textwidth]{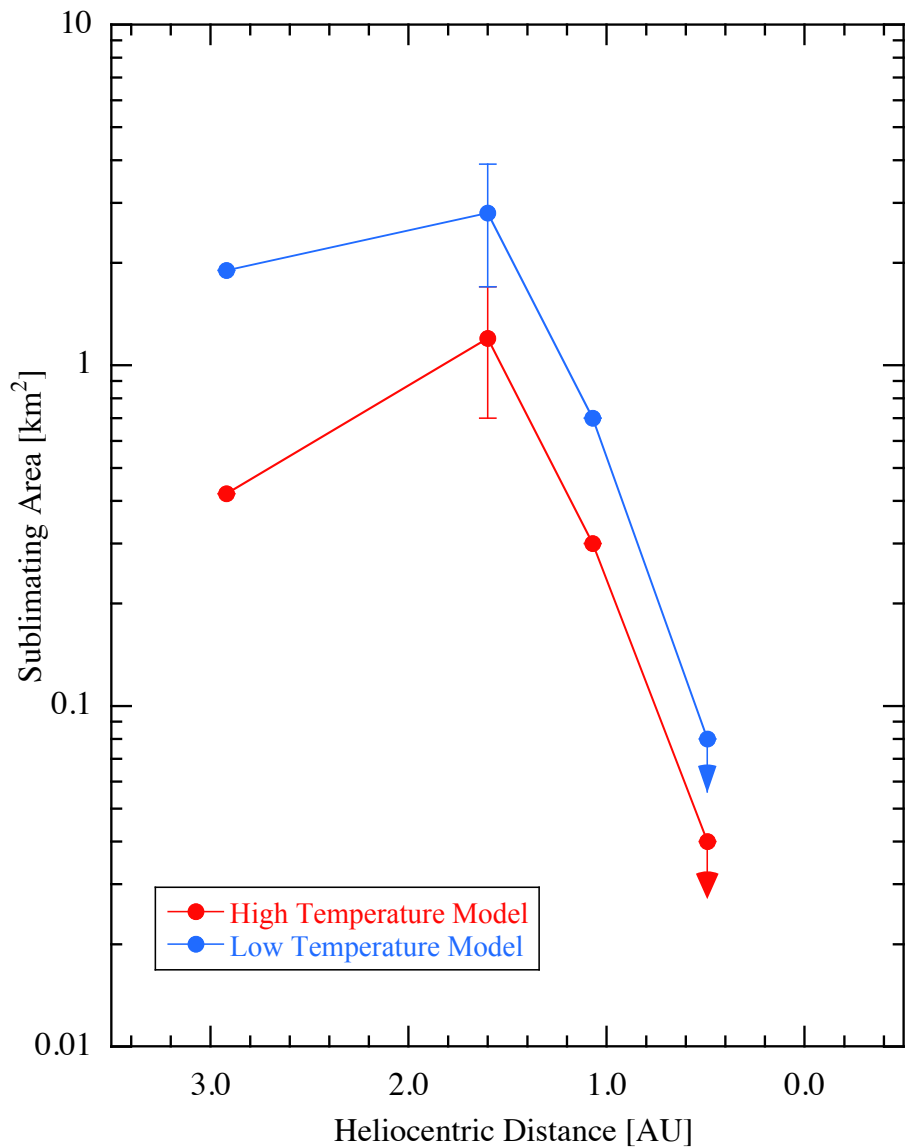}
\caption{Sublimating area as a function of heliocentric distance. Blue (red) points correspond to models in which the sublimation is assumed to occur at the lowest (highest) equilibrium temperatures for a given heliocentric distance.  Arrows signify upper limits based on non-detections of gaseous emission lines.  \label{a_vs_R} }
\end{center} 
\end{figure}

\noindent Values of $A_s$ computed from the gas emission observations and Equations (\ref{energy}) and  (\ref{A_s}) are listed in Table (\ref{areas}) and plotted against heliocentric distance in Figure (\ref{a_vs_R}).  In the Figure, red and blue points distinguish sublimating areas computed using the high ($\chi$ = 1) and low ($\chi$ = 2) temperature approximations, respectively.  Solutions between the red and blue lines are allowed by the data.  It is evident that the sublimating area decreased as Elenin approached the Sun, particularly inside 1 AU heliocentric distance.  This is different from the trend expected of a long-lived, un-evolving source (which would plot as a horizontal line in the Figure).  Beyond $r_H\sim$ 1 AU the gas production rates are compatible with sublimation from an area $A_s \sim$ 1 km$^2$.  

\subsection{The Nucleus}
The sublimating area, $A_s$, provides a useful but imperfect estimate of  the nucleus size.  With $A_s$ = 1 km$^2$, we compute the effective sublimation radius $r_s$ = ($A_s/\pi)^{1/2} \sim$ 0.6 km.  The estimate is imperfect because the nucleus could be larger than this but with a smaller fraction of its surface in sublimation. It is also possible that some fraction of the gas is produced by sublimation from  dust in the coma, rather than from the nucleus.   

Non-detections in the CFHT data set upper limits to the surviving  nucleus fragment sizes, through Equation (\ref{inversesq}).  On UT 2011 October 22,  the heliocentric and geocentric distances were $r_H$= 1.055 AU, $\Delta$ = 0.246 AU, respectively, and the phase angle was $\alpha$ = 69.3\degr.  For an object with solar colors, the Sloan $r'$ magnitude is related to the Johnson V magnitude by $r' =m_V  - 0.16$ \citep{2002AJ....123.2121S}, giving a limiting magnitude from the CFHT data of $m_V = 24.56$.  The phase function of possible nucleus fragments is unknown, but measurements of other cometary nuclei are broadly compatible with 0.04$\pm$0.01 magnitudes per degree \citep{2004come.book..223L}.  After correcting for the distances and phase angle, we find a limit to the absolute magnitude of surviving nucleus fragments of $m_V(1,1,0) \ge 24.72\pm0.7$. The large uncertainty on $m_V(1,1,0)$ reflects mainly the effect of the unknown phase function. 
We substitute into Equation (\ref{inversesq}) assuming $p_V$ = 0.04, as is typical of the nuclei of comets \citep{2004come.book..223L}, to find an upper limit to the equivalent circular area radius of any surviving nucleus fragment  as $r_{max} = (C_e/\pi)^{1/2} \le$ 40 m.   Larger fragments, unless of much lower albedo (or perhaps of much steeper phase function), would have been detected in the CFHT images.

\section{DISCUSSION}
\subsection{Nature of the Breakup}

We have obtained the following constraints on the breakup of the nucleus of Elenin: 

\begin{itemize}
\item A crude estimate of the radius of the nucleus before breakup is $r_n \sim$ 0.6 km, based on the measured gas production rates.  
\item The largest bodies remaining after breakup had radii $r_{max} \le$ 40 m.  
\item The break up event started approximately on UT 2011 August 18$\pm1$ at $r_H$ = 0.7 AU.
\item The peak cross-section of the debris produced by the breakup, as determined from the peak centered on August 30 at $m_R(1,1,0) = 9.5\pm0.2$ and Equation (\ref{inversesq}), was $C_e$ = 4300 $\pm$900 km$^2$.
\end{itemize}

We obtain a constraint on the size distribution of the post-fragmentation particles as follows.  

We suppose that the nucleus fragments into a differential power-law size distribution in which $n(a)da = \Gamma a^{-q} da$.  Here, $\Gamma$ and $q$ are constants of the distribution, and the particles span the size range from $a_{min}$ to $a_{max}$.  The combined cross-section of such a power-law distribution is

\begin{equation}
C_{e} = \int_{a_{min}}^{a_{max}} \pi \Gamma a^{2-q}  da.
\label{C}
\end{equation}

\noindent In a breakup, the total mass of the fragments must be equal to the initial mass of the nucleus.  We write 

\begin{equation}
\frac{4}{3} \pi \rho_n r_n^3 = \int_{a_{min}}^{a_{max}} \frac{4\pi \Gamma \rho}{3} a^{3-q}  da
\label{M}
\end{equation}

\noindent where $\rho_n$ and $r_n$ are the nucleus density and radius, respectively, and $\rho$ is the density of the ejected particles.  We combine Equations (\ref{C}) and  (\ref{M}) to eliminate $\Gamma$, obtaining

\begin{equation}
\rho_n r_n^3 = \frac{\rho C_e}{\pi} \frac{\int_{a_{min}}^{a_{max}}  a^{3-q}  da}{\int_{a_{min}}^{a_{max}}   a^{2-q}  da}.
\label{equality}
\end{equation}

\noindent Particles smaller than $\sim$10$^{-7}$ m are inefficient scatterers of optical photons and will contribute negligibly to the measured cross-section, $C_e$, while objects larger than 40 m should have been detected in the post-outburst CFHT images, but were not.  Accordingly, we set  $a_{min}$ = 10$^{-7}$ m and $a_{max}$ = 40 m, respectively.  We assume that $\rho$ = $\rho_n$, and then, given $a_{max} \gg a_{min}$, we approximate the solution to Equation (\ref{equality}) by

\begin{equation}
r_n^3 = \frac{ C_e}{\pi} \left(\frac{3-q}{4-q}\right) \frac{a_{max}^{4-q}}{a_{min}^{3-q}}
\label{qsolver}
\end{equation}

\noindent provided $q \neq 3, 4$.  We solved Equation (\ref{qsolver}) by Newton-Raphson iteration using the values of $C_e$ and $r_n$ found earlier, to find $q$ = 3.24. By substitution into Equation (\ref{C}), we obtain $\Gamma$ = 3.7$\times$10$^7$.   The derived value of $q$ is relatively insensitive to the assumed input parameters. For example, changing $r_n$ from 0.3 km to 1.2 km changes $q$ from 3.36 to 3.06.  Changing $a_{max}$, $a_{min}$ or $C_e$ even by an order of magnitude has a similar or smaller effect.  While we make no attempt to define a formal statistical uncertainty, from a range of experiments we are confident that distributions with $q$ = 3.3$\pm$0.2 encompass the likely range of input parameter uncertainties.

In a $q$ = 3.3 distribution, the mass is carried by the largest particles while the cross-section is dominated by small particles.  To see this, we calculate $a_{1/2}$, the particle radius below which the integrated cross-section is half the total cross-section, from

\begin{equation}
\frac{1}{2} = \frac{\int_{a_{min}}^{a_{1/2}} a^{2-q} da}{\int_{a_{min}}^{a_{max}} a^{2-q} da}.
\label{ahalf}
\end{equation}

\noindent With $q$ = 3.3 we solve Equation (\ref{ahalf}) to find $a_{1/2}$ = 1 $\mu m$.  Half the cross-section, and presumably half the scattered intensity, is carried by particles smaller than 1 $\mu m$.  However, these sub-micron particles carry only $\sim$5$\times$10$^{-6}$ of the total mass.

Small particles, in addition to having a strong effect on the scattered intensity, can be strongly affected by solar radiation pressure.  Particles accelerated by a constant radiation pressure travel a distance from their source given by

\begin{equation}
\delta x = \frac{1}{2}\beta g_{\odot} \delta t^2
\label{dx}
\end{equation}

\noindent where $\beta$ is the ratio of radiation pressure acceleration to gravitational acceleration, $g_{\odot}$, and $\delta t$ is the time elapsed since ejection. Particles having a wide range of shapes and compositions are approximately represented by $\beta = 1/a_{\mu m}$, where $a_{\mu m}$ is the particle radius expressed in microns \citep{boh83}.   At $r_H$= 0.7 AU, the solar gravity is $g_{\odot}$ = 0.012 m s$^{-2}$.  Expressing the distance $\delta x$ in kilometers and the time $\delta t$ in days, we write Equation (\ref{dx}) as

\begin{equation}
\delta x \sim \frac{45,000}{a_{\mu m}}~\delta t^2.
\label{distance2}
\end{equation}

\noindent In Figure (\ref{ntt}), the longest dimension of the debris field on UT 2011 August 28 is of order 100\arcsec,~corresponding to $\delta x \sim$60,000 km in the plane of the sky.  Equation (\ref{distance2}) indicates that 1 $\mu$m particles could have been accelerated over the full length of the debris cloud on timescales as short as $\delta t \sim$ 1 day.  The comet was pictured by \citet{2011CBET.2876....1S} on Oct. 21 and 23 ($\delta t $= 64 and 66 days) from the ground.  A faint fan-like cloud with an extended size 1\arcdeg.5$\times$10\arcmin.0 was seen in the expected position of the comet, corresponding to $\sim 10^6$ km in the plane of the sky  at the geocentric distance $\Delta \sim$ 0.25 AU.   This extension corresponds, by Equation (\ref{distance2}), to the movement of particles up to radius $a \sim 200 ~\mu m$ in the time since the start of the photometric outburst.  Of course, smaller particles could have travelled this distance in less time, if the outburst were not impulsive.

Larger particles, if moving only under the action of radiation pressure, should be confined closer to the location of the former nucleus.  For example, 10\% of the cross-section in the above $q$ = 3.3 distribution is carried by particles with $a > 200~\mu m$ (0.2 mm).  Such particles would be displaced by radiation pressure over a distance $\delta x \sim$ 23,000 km in the NTT image taken on UT August 28 ($\delta t$ = 10 days) according to Equation (\ref{distance2}).   By the same equation, only particles $a \gtrsim 12$ mm would remain within a distance $\delta x <350$ km 10 days after release, this being the approximate size of the seeing disk in the NTT observations of UT 2011 August 28.  Such particles carry $<$ 1\% of the total cross-section, explaining why the distended comet on this date displays no evidence for a strong central condensation.

\subsection{Mechanism of the Breakup}
\begin{figure}
\epsscale{0.9}
\begin{center}
\includegraphics[width=0.9\textwidth]{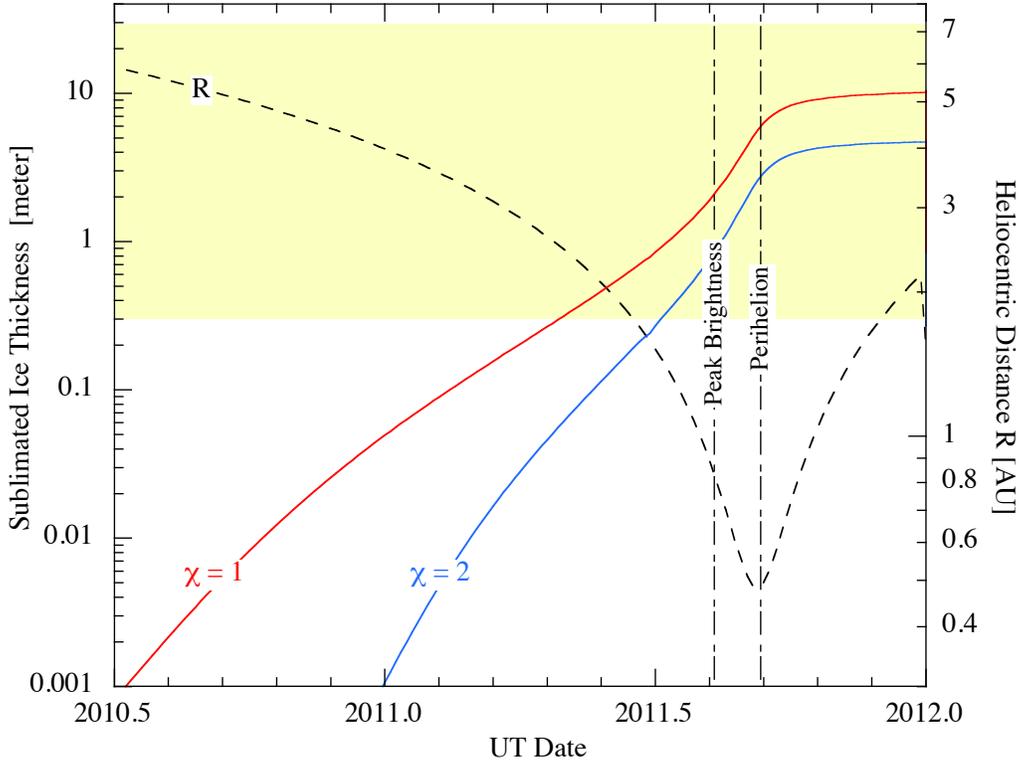}
\caption{Integrated thickness of ice layer lost by sublimation on Elenin as a function of the date. The red and blue curves correspond to the high and low temperature limits ($\chi$ = 1 and $\chi$ = 2, respectively) in Equation (\ref{ell}). The yellow shaded region marks the range of ice layer thicknesses which must be lost in order to change the nucleus angular momentum by a factor of order unity, according to Equation (\ref{torque}).  The black short-dashed line shows the heliocentric distance as a function of date, for reference (right hand axis).  The dates of peak brightness (UT 2011 August 12) and perihelion (UT 2011 September 10)  are marked.   \label{ice_loss} }
\end{center} 
\end{figure}

Could the nucleus of Elenin have simply sublimated away?  We used Equation (\ref{energy}) to estimate the thickness of ice that could be sublimated on the pre-perihelion leg of Elenin's orbit.  The equation was solved for $F_s$ as a function of heliocentric distance, $r_H$, and the equivalent thickness of sublimated ice was then calculated from 

\begin{equation}
\Delta r_n = \int_{t_0}^t \frac{F_s}{\rho_n}  dt
\label{ell}
\end{equation}

\noindent where $\rho_n$ = 1000 kg m$^{-3}$ is the nominal density.  We start the integration at time $t_0$ = 0, corresponding to Elenin at heliocentric distance 5 AU where water ice sublimation is negligible.  The result of Equations (\ref{energy}) and (\ref{ell}) is shown in Figure (\ref{ice_loss}), for the two limiting values of parameter $\chi$.  

Figure (\ref{ice_loss}) shows that sublimation losses on the journey inwards to perihelion are tiny (roughly $\Delta r_n =$ 2 to 5 m, for the two models) compared to the $r_n \sim$ 600 m radius of the nucleus. Therefore, sublimation losses cannot be directly responsible for the shrinkage and disappearance of the nucleus of Elenin, which requires $\Delta r_n / r_n$ = 1.  This is unlike the case of sun-grazing comets (i.e.~those which reach perihelion distances of a few solar radii). Many of these bodies have dimensions of order 10 meters \citep{2010AJ....139..926K}, and are small enough to simply sublimate away \citep{2011A&A...535A..71B}.  Comet Elenin also disintegrated too far from the Sun for tidal stresses to have played a role (the Roche radius for a nucleus of density 1000 kg m$^{-3}$ is $\sim 2.5 R_{\odot} \sim$ 0.01 AU).  Gas pressures produced by sublimation and ram pressure with the solar wind are likewise both utterly negligible at $r_H$ = 0.7 AU.  

It has been suggested that torques exerted by the loss of material from the nucleus can lead to rotational breakup, and that rotational disruption is the likely dominant mode of destruction of comets \citep{1992LIACo..30...85J,1997EM&P...79...35J}.  Observationally, \citet{1991ICQ....13...89B} has noted that intrinsically faint (presumably small nucleus) long-period comets with perihelia $<$0.5 AU vanish with higher probability than bright (presumably large nucleus) long-period comets.   Rotational breakup is especially effective for small cometary nuclei, suggesting that it may be responsible for the demise of Elenin and small long-period comets, generally. Accordingly, we examine the hypothesis that the disintegration of Elenin was precipitated by a rotational instability.  

Mass lost from a nucleus results in a change in the angular momentum, given by $\Delta L = k_T \Delta M v_{th} r_n$, where $\Delta M$ is the mass lost, $v_{th}$ is the outflow speed at which the mass leaves, and $r_n$ is the nucleus radius.  Dimensionless parameter $k_T$ is the ratio of the moment arm for the torque to the nucleus radius.  Simple estimates give $k_T \sim$ 0.05 \citep{1997EM&P...79...35J} while measurements from two comets give 0.005 $\le k_T \le$ 0.04 (for P/Tempel 1 \citep{2011Icar..213..345B}) and $k_T \sim$ 0.0004 (for 103P/Hartley 2 \citep{2011ApJ...734L...4D}).  We take $k_T \sim$ 10$^{-2\pm1}$, to reflect the dispersion in these values.  We identify $\Delta M$ with the mass of a spherical shell on the nucleus, $\Delta M = 4\pi r_n^2 \rho_n \Delta r_n$, where $\rho_n$ is the nucleus density and $\Delta r_n$ is the thickness of the shell.  Then, $\Delta L = 4 \pi k_T r_n^3 \rho_n v_{th} \Delta r_n$.  The angular momentum of the nucleus is $L = k_0 M_n r_n^2 \omega$, where $k_0$ = 2/5 for a homogeneous sphere of mass $M_n$ and $\omega = 2 \pi/P_0$ is the angular frequency of rotation at initial period $P_0$.  We substitute $M_n = 4\pi r_n^3 \rho_n/3$ and set $L = \Delta L$ to estimate the thickness of the layer which must be lost to modify the nucleus angular momentum by a factor of order unity;

\begin{equation}
\Delta r_n \sim \frac{2 \pi k_o r_n^2}{3 k_T v_{th} P_0}.
\label{torque}
\end{equation}

\noindent We substitute $v_{th}$ = 500 m s$^{-1}$ corresponding to the sound speed at 200 K, the approximate equilibrium temperature of ice freely sublimating at 1 AU.  We assume $P_0$ = 5 hr, typical of kilometer-sized small bodies, $k_T$ = 10$^{-2\pm 1}$ and $r_n$ = 600 m to find 0.3 $\le \Delta r_n \le$ 30 m.  This is very approximate, not least because several of the parameters in Equation (\ref{torque}) are unmeasured. Nevertheless, it is permissible to note that for most observationally allowed choices of the parameters, the equation gives $\Delta r_n / r_n \ll 1$, meaning that only a tiny fraction of the nucleus radius needs to be shed in order to change the angular momentum by a large factor.  For comparison, the destruction of the nucleus by sublimation alone requires $\Delta r_n / r_n = 1$, does not lead to a sudden break-up, and takes much longer than the time taken by Elenin to travel from its discovery distance to perihelion.

The range 0.3 $\le \Delta r_n \le$ 30 m is shown in yellow in Figure (\ref{ice_loss}).  The Figure shows that both the low and high temperature model curves, shown as blue and red lines, respectively, cross into the yellow shaded zone at pre-perihelion distances about 1.5 AU and 2.5 AU, respectively.  Both distances are reached before the date of peak brightness.   While this is far from proof that Elenin disintegrated because of its own outgassing torques, we must conclude that this is a plausible mechanism worthy of further consideration.

\clearpage

\section{SUMMARY}
We present a study of long-period comet C/2010 X1 (Elenin), using observations taken with a variety of telescopes over a wide range of heliocentric distances and phase angles.  We find that:

\begin{enumerate}

\item The comet displayed a surge in apparent brightness by $\sim$11 magnitudes centered on UT 2011 August 12.95 (DOY 224.95), when pre-perihelion at heliocentric distance $r_H$= 0.84 AU.  The peak of this brightness surge is coincident with the passage of the comet through maximum phase angle ($\alpha = $172\degr.6~on DOY 225), showing that it is caused by strong forward-scattering by cometary dust, not by outburst or nucleus disintegration.

\item A previously undetected, much smaller brightness surge of $\sim$3 magnitudes occurred on UT 2011 August 30 (DOY 242), about 17 days after the forward-scattering peak.  This peak is unrelated to the phase angle and instead represents the true breakup of the nucleus, which began at heliocentric distance $r_H$ = 0.74 AU on UT 2011 August 18$\pm$1 (DOY 230$\pm$1), some 23$\pm$1 days before perihelion on UT September 10 (DOY 253). The dust cross-section is estimated 4300$\pm$900 km$^2$ at the peak of the brightness surge.

\item The measured gas production rates are consistent with equilibrium sublimation from an area $\sim$1 km$^2$, providing a crude estimate of the pre-disruption radius of the nucleus $r_n \sim$ 0.6 km.   Deep images after the breakup set an upper limit to radius of surviving fragments near 40 m (geometric albedo 0.04 assumed).  

\item The increase in the cross-section implied by the brightening is consistent with breakup of a 0.6 km radius spherical nucleus into a power-law distribution of fragment sizes.  We find a power-law index $q$ = 3.3$\pm$0.2, such that half of the scattering cross-section lies within sub-micron dust particles while most of the mass is in large particles, up to the limiting size established by CFHT observations.

\item The pre-perihelion disintegration of Comet Elenin starting at about 0.7 AU cannot be explained by simple sublimation losses, by gravitational stresses or by sublimation stresses.  We propose that the nucleus was accelerated to rotational breakup by torques imposed by sublimation-driven mass-loss.

\end{enumerate}

\acknowledgements
We thank Bernard H\"{a}usler  and Pedro Lacerda for attempting post-perihelion observations on our behalf, and Rachel Stevenson and Jan Kleyna for help with SWARP.  We thank Bin Yang, Masateru Ishiguro, and Man-To Hui for reading the manuscript and providing comments. We thank the referee, Mathew Knight, for his speedy review and valuable comments. We thank the CFHT Director for kindly allocating discretionary time at short notice for this work.  
The Heliospheric Imager instrument was developed by a collaboration that included the University of Birmingham and the Rutherford Appleton Laboratory, both in the UK, the Centre Spatial de Liege (CSL), Belgium, and the U.S. Naval Research Laboratory (NRL), Washington DC, USA. The STEREO/SECCHI project is an international collaboration.  This work was supported, in part, by a grant to DJ from NASA's Origins program.
\clearpage 

\begin{deluxetable}{llcll}
\tablenum{1}
\tabletypesize{\scriptsize}
\tablecaption{Observations in 2011}
\label{elenin}
\tablewidth{0pt}
\tablehead{
\colhead{Dates [UT]} &
\colhead{Telescope} &
\colhead{Field of View} &
\colhead{Pixel Size} &
\colhead{Wavelengths} \\
 }
\startdata
Feb 10 - Jun 03 & KAIT& 6\arcmin.8& 0.8\arcsec & R-filter \\
May 16 - Jun 20 & STEREO  HI-1A & 20\arcdeg & 70\arcsec & 630-730 nm\\
Aug 01 - 05\tablenotemark{\dag} & STEREO HI-2 B & 70\arcdeg & 240\arcsec & 400-1000 nm\\
 Aug 06 - 12\tablenotemark{\dag} & STEREO HI-1 B  & 20\arcdeg & 70\arcsec & 630-730 nm\\
Aug. 14 - Sep. 04 & STEREO HI-1 B &  20\arcdeg & 70\arcsec & 630-730 nm\\
July 23, August 28 & NTT &  4.1\arcmin& 0.24\arcsec & R642 nm \\
Sep 01 - 22 & STEREO HI-1 A &  20\arcdeg & 70\arcsec & 630-730 nm\\
Oct 22 & CFHT & 1\degr & 0.187\arcsec & R-filter \\
Nov 11 & CFHT & 1\degr & 0.187\arcsec & R-filter \\
\enddata
\tablenotetext{\dag}{During these time periods, STEREO B was specially maneuvered to be rolled toward the comet.}
\end{deluxetable}
\clearpage

\clearpage

\begin{deluxetable}{lrclllrrl}
\tablenum{2}
\tabletypesize{\scriptsize}
\tablecaption{Gas Production Rates and Active Areas}
\label{areas}
\tablewidth{0pt}
\tablehead{
\colhead{UT Date} &
\colhead{$Q_{}$\tablenotemark{a}} &
\colhead{$dM_{}/dt$\tablenotemark{b}} &
\colhead{$r_H$[AU]\tablenotemark{c}} &
\colhead{$F_s(min)$\tablenotemark{d}} &
\colhead{$F_s(max)$\tablenotemark{e}} &
\colhead{$A_s(min)$\tablenotemark{f}}&
\colhead{$A_s(max)$\tablenotemark{g}}&
\colhead{Reference}
}
\startdata
2011 Mar 28 & 4.5$\times$10$^{26}$ & 12 & 2.92 & 6.4$\times$10$^{-6}$ & 3.0$\times$10$^{-5}$ & 0.4 & 1.9 & Korsun et al.~(2012)\\
2011 Jun 28 & 6.3$\pm$2.5$\times$10$^{27}$ & 190$\pm$80 & 1.60 & 6.6$\times$10$^{-5}$ & 1.6$\times$10$^{-4}$ & 1.2$\pm$0.5 & 2.8$\pm$1.1 & Schleicher (2015)\\
2011 Jul 30 & 4.8$\times$10$^{27}$ &  144 & 1.07 & 1.8$\times$10$^{-4}$ & 3.9$\times$10$^{-4}$ & 0.4 & 0.8 & Drahus et al.~(2011)\\
2011 Sep 07 & $<$3$\times$10$^{27}$ & $<$90 & 0.49 & 9.8$\times$10$^{-4}$ & 2.0$\times$10$^{-3}$ & $<$0.05 & $<$0.09 & Lovell et al.~(2011)\\
\enddata
\tablenotetext{a}{Estimated H$_2$O production rates, molecules s$^{-1}$}
\tablenotetext{b}{Equivalent H$_2$O gas mass production rate, kg s$^{-1}$}
\tablenotetext{c}{Heliocentric distance, AU} 
\tablenotetext{d}{Minimum equilibrium sublimation flux from Equation (\ref{energy}), kg m$^{-2}$ s$^{-1}$ } 
\tablenotetext{e}{Maximum equilibrium sublimation flux from Equation (\ref{energy}), kg m$^{-2}$ s$^{-1}$} 
\tablenotetext{f}{Minimum exposed ice area from Equation (\ref{A_s}), km$^2$} 
\tablenotetext{g}{Maximum exposed ice area from Equation (\ref{A_s}), km$^2$} 
\end{deluxetable}

\clearpage

\clearpage

\clearpage

\clearpage
\clearpage

\clearpage

\clearpage

\clearpage

\clearpage

\clearpage

\clearpage

\clearpage

\clearpage

\end{document}